\documentclass[11pt]{article}

\usepackage{amsfonts,latexsym,eucal,color,graphicx,epsfig,amssymb,amsmath,mathrsfs}
\usepackage{doi}

\usepackage{hyperref}

\numberwithin{equation}{section}

\newcommand{\be}{\begin{eqnarray}}
\newcommand{\ee}{\end{eqnarray}}
\newcommand{\beq}{\begin{eqnarray}}
\newcommand{\eeq}{\end{eqnarray}}

\newcommand{\dalm}{\kern1pt\vbox{\hrule height 0.9pt\hbox{\vrule width 0.9pt\hskip 2.5pt\vbox{\vskip 5.5pt}\hskip 3pt\vrule width 0.3pt}\hrule height 0.3pt}\kern1pt}

\newcommand{\tr}{\textrm{tr}}

\newcommand{\zt}{\zeta}	
\newcommand{\bzt}{\bar{\zeta}}

\newcommand{\bz}{\bar{z}}
\begin{document}


\begin{flushright}
DAMTP-2016-40
\end{flushright}
\vspace{.5cm}
\begin{center}{\Huge \bf Gravity Duals of Boundary Cones}
\vspace{1.5cm}

{\Large Joan Camps}
\vspace{.5cm}

{\it DAMTP, Cambridge University

Wilberforce Road, CB3 0WA Cambridge, United Kingdom}
\vspace{.5cm}

{\verb"j.camps@damtp.cam.ac.uk"}
\vspace{2cm}
\end{center}
\abstract{The replica trick defines R\'enyi entropies as partition functions on conically singular geometries. We discuss their gravity duals: regular bulk solutions to the Einstein equations inducing conically singular metrics at the boundary. When the conical singularity is supported on a flat or spherical surface, these solutions are rewritings of the hyperbolic black hole. For more general shapes, these solutions are new. We construct them perturbatively in a double expansion in the distance and strength of the conical singularity, and extract the vacuum polarisation due to the cone. Recent results about the structure of logarithmic divergences of R\'enyi entropies are reproduced ---in particular, $f_b\neq f_c$. We discuss in detail the dynamical resolution of the singularity in the bulk. This resolution is in agreement with a previous proposal, and indicates a non-minimal settling to the `splitting problem': an apparent ambiguity in the holographic entropy formula of certain theories with higher derivatives.}
\newpage
\tableofcontents
\newpage

\section{Introduction and summary}
The Ryu-Takayanagi area formula \cite{Ryu:2006bv, Ryu:2006ef} seeds many recent insights from quantum information into quantum gravity. This formula calculates the entanglement entropy of a field theory region as the area of a minimal surface in the gravity dual, generalising the Bekenstein-Hawking entropy.

Building on work by Casini, Huerta and Myers \cite{Casini:2011kv}, Lewkowycz and Maldacena have constructed a derivation of the Ryu-Takayanagi formula. This derivation, called `generalized entropy' \cite{Lewkowycz:2013nqa}, uses the replica trick in field theory and its gravity dual, which is in the Euclidean quantum gravity regime.

In this context, the replica trick defines entanglement entropy $S$ as the $n\rightarrow 1$ limit of R\'enyi entropies $S_n$. Using Euclidean techniques, the  $S_n$ can be related to field theory partition functions on conically singular manifolds \cite{Callan:1994py}. The conical singularity has support on the entangling surface that bounds the partition whose entanglement is being calculated. The period of the cycle contracting on the entangling surface is $2\pi n$, and the conical singularity is absent when $n=1$.

In a CFT, conformal symmetry can be used to send this entangling surface to infinity. Doing so leaves behind an Euclidean field theory geometry with a non-contractible cycle. This makes the setup analogous to that of thermal field theory. When these thermal states have black holes as gravity duals, arguments from Euclidean quantum gravity apply. An area law emerges as a Gibbons-Hawking derivation of black hole entropy \cite{Gibbons:1976ue} ---in which regularity of the bulk Euclidean geometry for all values of the temperature (the period of the thermal cycle) plays a central role.

\cite{Casini:2011kv} implemented these ideas for a spherical entangling surface in the vacuum of a CFT in flat space. This conformally maps to the Euclidean thermal hyperboloid, whose gravity dual is well known ---it is the hyperbolic black hole \cite{Emparan:1999gf}. Entanglement entropy in the flat space picture is related to thermal entropy in the hyperboloid, in which case it is dual to black hole entropy. Using Bekenstein-Hawking for the thermal case, a Ryu-Takayanagi area follows for the entangling one.

\cite{Lewkowycz:2013nqa} extended this picture to more general states and entangling regions. This involves field theory geometries for which the entangling cycle no longer generates a symmetry, and thus the connection to thermal physics weakens. The subtleties of the replica trick become more prominent, and the analytic continuation of the dual geometries to non-integer values of $n$, relevant for the $n\rightarrow 1$ limit, is less direct and needs to be discussed in detail.

Both \cite{Casini:2011kv} and \cite{Lewkowycz:2013nqa} work in `hyperbolic frames', in which the boundary conical singularity has been mapped to infinity. The primary goal of this paper is to carry out explicitly the construction in the `entangling frame', in which case the conical singularity of the boundary at $n\neq 1$ remains within sight, but nevertheless has a regular gravity dual. We will exhibit in detail these geometries, and discuss how exactly gravity in the bulk dynamically regulates boundary conical singularities.\footnote{The fact that such bulk geometries are regular was emphasised in \cite{Headrick:2010zt} and \cite{Hung:2011nu}.}

The new geometries are regular Euclidean solutions of the Einstein equations with a negative cosmological constant, subject to the boundary condition that the geometry induced at the conformal boundary has a conical singularity on a specified surface. To make analytic progress, we will find these geometries perturbatively in a double expansion in the distance and strength of the singularity, $(n-1)$. The distance is measured in units of the smallest lengthscale characterising the geometry of the background and surface supporting the singularity. We focus on the case of five bulk dimensions for convenience. An extension to general dimensions would be very interesting.

These geometries encode the vacuum polarisation of holographic conformal field theory due to conical singularities on general surfaces. Although strictly speaking there is no Fefferman-Graham expansion because the boundary metric is singular, we will extract an expectation value for the stress tensor. In our approximations, and ignoring contact terms, the stress tensor is traceless.

The Euclidean action of these bulk geometries relates to R\'enyi entropy $S_n$, that we extract to first order in $(n-1)$. We find that, contrary to the entanglement entropy term $S_1$, the first order term does not have an area law. Also, in precise agreement with \cite{Dong:2016wcf}, the logarithmic divergence of the first order term turns out to have a different structure from the entanglement entropy one.

The specific way in which gravity regulates these singularities impacts the generalisation of the Ryu-Takayanagi formula to theories with higher derivatives. This formula extends Wald's black hole entropy to setups without $U(1)$ symmetry. For theories without explicit derivatives of the Riemann tensor in their lagrangian,\footnote{see \cite{Miao:2014nxa} for a more general case.} it takes the schematic form \cite{Dong:2013qoa, Camps:2013zua}:
\beq
S=\int \frac{\partial L}{\partial \textrm{Riem}}+
\int\sum_{\alpha}\left(\frac{\partial^2 L}{\partial \textrm{Riem}^2}\right)_{\alpha}\frac{K^2}{1+q_{\alpha}} \,,
\eeq
where $L$ is the gravity lagrangian, $K$ is the extrinsic curvature, and $q_\alpha$ are coefficients characterising how exactly the conical singularity is regulated in the bulk (this is reviewed in sec~\ref{sec:Splitting}). We will see below that they differ from a minimal prescription.

The rest of this paper is organised as follows. Sec.~\ref{sec:HyperboloidsAndCones} reviews the Casini-Huerta-Myers construction and exhibits that Euclidean hyperbolic black holes can be written as smooth gravity duals to straight conical singularities. In the rest of the paper we will deform this cone away from straightness and explore the consequences of its gravity dual. Sec.~\ref{sec:FermiCoords} reviews the construction of Fermi-like coordinates adapted to codimension two surfaces, and the natural implementation of the replica trick in these coordinates. Sec.~\ref{sec:Solutions} constructs explicitly the gravity duals, and is the core of the paper. There are many ways to deform a surface away from straightness (equivalently, many ways to squash a cone \cite{Dowker:1994bj}), and, after a general overview, we proceed in a casuistic way. To the order of Riemann curvature, this results in thirteen subsections analysing different such deformations.\footnote{We take some advantage of conformal invariance in the boundary to reduce the number of cases from $2+18$ to $1+13$. Secs.~\ref{sec:Renyi} and \ref{sec:Splitting} only use the results of subsections \ref{sec:Kztztz} and \ref{sec:Rztbztztbzt}-\ref{sec:KztztzKbztbztbz}, and some readers may want to focus on these.} Sec.~\ref{sec:VacPol} summarises the results regarding the vacuum polarisation induced by these singularities, and sec.~\ref{sec:Renyi} explains how to reproduce the results of \cite{Dong:2016wcf} regarding logarithmic divergences of R\'enyi entropies in CFT. In sec.~\ref{sec:Splitting} we discuss the consequences of sec.~\ref{sec:Solutions} for the entropy formula of theories of gravity with higher-derivative interactions. We conclude in sec.~\ref{sec:Conclusions}.

\section{Hyperbolic black holes and boundary cones}\label{sec:HyperboloidsAndCones}
This section reviews some aspects of the Casini-Huerta-Myers construction \cite{Casini:2011kv} and hyperbolic black holes. It also serves to set notation and discuss coordinates that we will be using throughout.

Consider the R\'enyi entropy of a 4D conformal field theory in the vacuum across a straight plane.\footnote{Given a density matrix $\rho$ ---for instance, constructed by restricting the vacuum to a subset of degrees of freedom $A$, $\rho=\tr_{\bar{A}}|0\rangle\langle0|$---, its R\'enyi entropy is defined:
$$
S_n=\frac{1}{1-n}\log\left(\tr \rho^n\right)\,.
$$
} The replica trick maps this quantity to the Euclidean partition function on the conically singular geometry:
\beq
ds^2=r^2d\tau^2+dr^2+d\zt\, d\bzt\,,\qquad \tau\sim \tau+2\pi n\,,
\label{eq:ConePolar}\eeq
where we took complex coordinates  $\zt = \sigma^1+i\sigma^2$ on the entangling plane, at $r=0$, and the geometry is singular for $n\neq 1$ because of the period of $\tau$.

A convenient way to write cone metrics uses complex coordinates also in the plane of the cone. Define $z=r^{1/n} e^{i\tau/n}$. This is a good complex coordinate when $\tau$ has the period in \eqref{eq:ConePolar}, and the cone becomes
\begin{align}
ds^2=&(z\bz)^{n-1}n^2dz\, d\bz+d\zt\, d\bzt\,,
\label{eq:ConeComplex}
\end{align}
which can be obtained from Euclidean space by the `quotient' $z\rightarrow z^{n}$.

This conical geometry \eqref{eq:ConePolar} is conformal to the Euclidean thermal hyperboloid $S^1\times\mathbb{H}_3 $:
\beq
ds^2=d\tau^2+\frac{dr^2+d\zt\, d\bzt}{r^2}\,,\qquad \tau\sim\tau+2\pi n\,,
\label{eq:ThH}\eeq
which is regular for all periods of $\tau$.

If the CFT has a GR gravity dual, the geometry dual to the partition function on \eqref{eq:ThH} is the hyperbolic black hole \cite{Emparan:1999gf}:\footnote{Except in selected places, we work in units of the AdS radius, $\ell=1$.}
\beq
ds^2=\frac{d\rho^2}{f(\rho)}+f(\rho)d\tau^2+\rho^2\frac{dr^2+d\zt\, d\bzt}{r^2}\,,
\qquad f(\rho)=\rho^2-1-\frac{\rho_h^2(\rho_h^2-1)}{\rho^2}\,,
\label{eq:HypBH}\eeq
where
\beq
\rho_h=\frac{1+\sqrt{1+8n^2}}{4n}=1-\frac{n-1}{3}+O\left((n-1)^2\right)\,.
\eeq
The metric at the boundary, at $\rho\rightarrow\infty$, is \eqref{eq:ThH}. $\tau$ closes smoothly at $\rho=\rho_h$ when $\tau\sim\tau+2\pi n$. Hence this is a good holographic dual to \eqref{eq:ThH}. The geometry becomes AdS when $n=1$, when the temperature is $1/2\pi$ in units of the radius of the hyperboloid.

Boundary conformal transformations are implemented by large diffeomorphisms in the bulk, so there is a change of coordinates that writes the geometry \eqref{eq:HypBH} as the gravity dual of the conical singularity \eqref{eq:ConeComplex}, \cite{Hung:2011nu}. To leading order in $(n-1)$, one such diffeomorphism is:
\begin{align}
\rho&\rightarrow\sqrt{1+x\left(1+\frac{1}{x}\right)^{\frac{n-1}{n}}}\left(1-\frac{1}{3}\frac{(n-1)}{1+x}\right)
+O\left((n-1)^2\right)\nonumber\\
r&\rightarrow\rho\sqrt{1+x\left(1+\frac{1}{x}\right)^{\frac{n-1}{n}}}\left(1-\frac{1}{2}\frac{(n-1)}{1+x}\right)+O\left((n-1)^2\right)\nonumber\\
\tau&\rightarrow i\frac{n}{2}\log\frac{\bz}{z}\,,
\end{align}
where
\beq
x\equiv \frac{(z\bz)^n}{\rho^2}\,.
\label{eq:xdef}\eeq
Now \eqref{eq:HypBH} reads:
\begin{align}
ds^2=%
&\left(1-\frac{2}{3}\frac{(n-1)}{1+x}\right)\frac{d\rho^2}{\rho^2}
+ \left(1-\frac{2}{3}\frac{(n-1)}{1+x}\right)\frac{n^2\left({1+x}\right)^{\frac{n-1}{n}}dz\, d\bz}{\rho^{2/n}}
\nonumber\\
&+\left(1+\frac{1}{3}\frac{(n-1)}{1+x}\right)\frac{d\zt\, d\bzt}{\rho^2}
+O\left((n-1)^2\right)\,.
\label{eq:DualCone}
\end{align}

In these coordinates the boundary is at $\rho\rightarrow 0$. Its geometry is indeed conically singular, because for $\rho\rightarrow 0$:
\beq
g_{z\bz}\rightarrow  \frac{1}{2}\frac{(z\bz)^{n-1}n^2}{\rho^2}\,,
\eeq
and we recover \eqref{eq:ConeComplex} as the boundary metric. The axis of the cone extends into the bulk in a regular manner: at fixed $\rho$, as $z\bz\rightarrow 0$,
\beq
g_{z\bz}\rightarrow \frac{1}{2}\left(1-\frac{2}{3}(n-1)\right)\frac{n^2}{\rho^{2/n}}\,,
\label{eq:gzzbaxis}\eeq
a constant, and thus regular. The factors of $\frac{(n-1)}{1+x}$ are bounded corrections that remain small everywhere for small $(n-1)$. At fixed $\rho$, eq.~\eqref{eq:DualCone} is a regularised version of the boundary conical singularity, to which it tends far from the axis, $x\gg1$. Therefore, eq.~\eqref{eq:DualCone} is a boundary conical singularity that gravity dynamically regularises in the bulk.

\section{Fermi coordinates for field theory replicas}\label{sec:FermiCoords}
Following the program of generalized entropy \cite{Lewkowycz:2013nqa}, we will find the gravity duals to cones on entangling surfaces and backgrounds more general than the plane in flat space.

One can take coordinates adapted to a generic surface by shooting geodesics orthogonal to it. In an expansion near such surface the metric can be written as:
\begin{align}
ds^2=&\left(\gamma_{ij}+\left[2\kappa\, K_{ijz} z+\kappa^2Q_{ijzz}z^{2}+\kappa^2Q_{ijz\bz}z\bz+\textrm{c.c.}\right]\right)d\sigma^i d\sigma^j+2\kappa\, A_{iz\bz} (\bz\, dz-z\,d\bz)\,d\sigma^i \nonumber\\
&-\frac{4}{3}\kappa^2\left[R_{izz\bz}z-\textrm{c.c.}\right] (\bz\, dz-z\, d\bz) d\sigma^i+\left(1+2\kappa^2R_{z\bz z\bz}z\bz\right)dz\,d\bz+O(\kappa^3)\,,
\label{eq:AdaptedCoords}
\end{align}
where, as above, we parametrise the transverse geodesics in complex coordinates: $z$, $\bz$. The objects $\gamma_{ij}$, $K_{ijz}$, $Q_{ijz\bz}$, etc., characterise the embedding and background geometry on the surface and may depend on its coordinates $\sigma^i$. c.c.~stands for complex conjugation within the square brackets. $\kappa$ is a small book-keeping parameter counting powers of the distance to the surface in units of the characteristic lengthscale of the geometry. As in the previous section, it is convenient to take complex coordinates also in the surface and analogously expand around their origin. By appropriately constructing these coordinates, one can eliminate $A_{iz\bz}$ and its symmetrised first derivatives at the origin, as well as the Christoffel symbols of the induced metric $\gamma_{ij}$:
\begin{align}
\gamma_{ij}d\sigma^id\sigma^j &=  d\zt\,d\bzt-\frac{1}{3}\kappa^2{\cal R}_{\zt\bzt\zt\bzt}\left(\bzt\, d\zt-\zt\, d\bzt\right)^2
+O(\kappa^3)\,,\nonumber\\
2\kappa\, A_{iz\bz}d\sigma^i &=  -\kappa^2 F_{\zt \bzt z\bz}(\bzt\, d\zt - \zt\, d\bzt)+O(\kappa^3)\,,
\label{eq:SurfaceGauge}\end{align}
where $\kappa$ now also keeps track of the distance to the origin in the surface, $\zt\bzt=0$. $R_{\mu\nu\rho\sigma}$ denotes components of the background Riemann tensor, and ${\cal R}_{ijkl}$ refers to the Riemann tensor of $\gamma_{ij}$ ---the intrinsic curvature of the $z\bz=0$ surface.

A convenient feature of complex coordinates is that trace and traceless elements of symmetric tensors are readily distinguished. For example, $K_{\zt\bzt z}$ belongs to the trace of $K_{ijz}$, whereas $K_{\zt\zt z}$ belongs to its traceless part.

The use of these coordinates on entanglement entropy calculations in field theory was pioneered in \cite{Rosenhaus:2014woa, Rosenhaus:2014zza}, and \cite{Lewkowycz:2014jia} for R\'enyi entropy. We expand to $O(\kappa^2)$ because we are interested in effects due to Riemann curvature.
The Riemann tensor of \eqref{eq:AdaptedCoords}--\eqref{eq:SurfaceGauge} at $\zt\bzt=z\bz=0$ is fully captured at this order, and reads (setting $\kappa=1$):
\begin{align}
R_{ij}{}^{kl} &= {\cal R}_{ij}{}^{kl} - 4 K_{i}{}^{[k}{}_{z} K^{l]}{}_{j \bz} - 4  K_{i}{}^{[k}{}_{\bz} K^{l]}{}_{j z} \cr
R_{ij k}{}^{\bz} &=  2 \partial_{[i} K_{j] k}{}^{\bz} \cr
R_{ij}{}^{z\bz} &= F_{ij}{}^{z\bz} - 2 K_{[i}{}^{kz} K_{j]k}{}^{\bz} \cr
R_i{}^z{}_j{}^{\bz} &= \frac{1}{2} F_{ij}{}^{z\bz}- Q_{ij}{}^{z\bz}+K_{i}{}^{k\bz} K_{jk}{}^{z}\cr
R_{izjz} &=K_{i}{}^{k}{}_{z} K_{jkz} - Q_{ijzz}\,,
\label{eq:GC}\end{align}
$R_{iz z\bz}$ and $R_{z\bz z\bz}$ coming directly from \eqref{eq:AdaptedCoords}.

Only some combinations of the above objects transform covariantly under conformal transformations. When working with CFTs, one may take advantage of such symmetries to eliminate  non-covariant elements. For example, one can choose to eliminate the trace of the extrinsic curvature $K_{\zt\bzt z}$, as well as the intrinsic Riemann ${\cal R}_{\zt\bzt\zt\bzt}$, and the traces $Q_{\zt\bzt z\bz}$ and $Q_{\zt\bzt zz}$. For reasons that will be clear, we will drop $K_{\zt\bzt z}$ but will keep the ${\cal R}$ and $Q$s to keep track of the topology of the surface (via Gauss-Bonnet) and as a  device to check the conformal covariance of our results. One check of such covariance will be the appearance of the  trace of the projection of the bulk Weyl on the entangling surface:
\beq
W_{ij}{}^{ij}=\gamma^{\mu\rho}\gamma^{\nu\sigma}W_{\mu\nu\rho\sigma}=
\frac{8}{3}\left(2\,Q_{\zt\bzt z\bz}-R_{z\bz z\bz}-{\cal R}_{\zt\bzt\zt\bzt}\right)\,.
\label{eq:PWeyl2}\eeq

Now, as earlier, replicating around the $z\bz=0$ surface is implemented by $z\rightarrow z^n$. We will then be after the gravity duals of partition functions of CFTs on
\begin{align}
ds^2=&\left(\gamma_{ij}+\left[2\kappa\, K_{ijz}{} z^n+\kappa^2Q_{ijzz}z^{2n}+\kappa^2Q_{ijz\bz}(z\bz)^n+\textrm{c.c.}\right]\right)d\sigma^i d\sigma^j\nonumber\\
&+\left(2\kappa\, A_{iz\bz}
-\frac{4}{3}\kappa^2\left[R_{izz\bz}z^n-\textrm{c.c.}\right]\right)n(z\bz)^{n-1} (\bz\, dz-z\, d\bz) d\sigma^i\nonumber\\
&+\left(1+2\kappa^2R_{z\bz z\bz}(z\bz)^n\right)n^2(z\bz)^{n-1}dz\,d\bz+O(\kappa^3)\,,
\label{eq:AdaptedRepl}
\end{align}
with $\gamma_{ij}$ and $A_{iz\bz}$ as in \eqref{eq:SurfaceGauge}.

$z^n$ is multivalued for generic $n$, and so \eqref{eq:AdaptedRepl} is a geometry only for $n$ a positive integer (in particular, only then $ds^2$ is continuous for $z$ in the complex plane). We will nevertheless treat $n$ as a real number, expand $ds^2$ in powers of $(n-1)$, and speak about a CFT in \eqref{eq:AdaptedRepl} for real $n$. This is tantamount to speaking about a CFT on an `analytic continuation of a geometry', and it is an abuse of language. Quantities of interest for $n\in \mathbb{R}$ should be thought as being analytically continued from $n\in\mathbb{Z}$, as R\'enyi entropies in the replica trick. The usefulness of this picture is that it picks the right analytic continuation for the quantities of interest \cite{Lewkowycz:2013nqa}.

\section{Gravity duals of squashed conical singularities}\label{sec:Solutions}
This section is the core of the paper, where the regular gravity duals to \eqref{eq:AdaptedRepl} are spelled out, to first order in $(n-1)$ and to second order in $\kappa$. We start by giving an overall picture of these geometries and their properties.

The expansion in $\kappa$ is a derivative expansion, and for this reason our results may be reminiscent of other such expansions in gravity, as fluids/gravity \cite{Bhattacharyya:2008jc, Bhattacharyya:2008mz} or blackfolds \cite{Emparan:2009at, Camps:2012hw}. The strategy is to deform the boundary metric in \eqref{eq:DualCone} in the expansion of \eqref{eq:AdaptedRepl}. This generically does not solve the Einstein equations by itself, and one needs to add a small correction to that end. We then solve for this correction subject to the boundary conditions of normalisability and regularity.

The notion of bulk regularity we are alluding to at non-integer $n$ is not a standard one if there is squashing. The metrics we will call `regular' have $1/r$ divergent curvature invariants at the bulk axis,\footnote{These singular invariants are on top of the multivaluedness discussed above, that extends into the bulk. In the presence of replica symmetry, multivaluedness can be eliminated by a quotient by that symmetry. This returns a regular boundary and a conically singular bulk, that still has the same $1/r$ divergent curvature invariants.} but \emph{(i)} these divergences drop from the field equations, and \emph{(ii)} they are altogether absent at integer $n$. Condition \emph{(ii)} is the usual AdS/CFT bulk regularity at integer $n$, while \emph{(i)} picks the right analytic continuation of the bulk metric to non-integer $n$ \cite{Lewkowycz:2013nqa}. In a sense, this is the closest we can get to the usual notion of regularity for $n\in\mathbb{R}$.

We will assume that all the non-trivial dependence of the corrections is in the dimensionless $x$ of eq.~\eqref{eq:xdef}. The geometry then does not have any more dependence on the angle in which we approach the singularity other than the one following from the index structures in \eqref{eq:AdaptedRepl}. This reduces the equations for the corrections to ODEs. To leading order in $(n-1)$, our notion of regularity at the bulk axis, $x=0$, boils down to the expandability of the metric in non-negative powers of $x^{\frac{1}{n}}$ and $x^{\frac{n-1}{n}}$ \cite{Camps:2014voa};\footnote{that is, expandability in positive powers of $z\bz$ and $(z\bz)^{n-1}$ near the bulk axis, at $z=\bz=0$ and finite $\rho$.} and that for $g_{z\bz}$ such  expansion has a constant term, \eqref{eq:gzzbaxis}. This gives smoothness at positive $n\in\mathbb{Z}$. Assuming dependence just in $x$ implies replica symmetry ---a discrete rotational symmetry in the plane of the cone, $z\rightarrow z\, e^{ik/n}$ for $k\in\mathbb{Z}_n$.\footnote{Since we construct them by $z\rightarrow z^n$, the boundary metrics are automatically replica symmetric, but the bulk could break the symmetry spontaneously \cite{Camps:2014voa}.}

This dependence on $x$ implies covariance under diffeomorphisms in the surface, that we will maintain explicitly. This is a useful principle when writing ansatze for the bulk corrections, because it forbids appearances of the surface coordinates $\zt$ other than the ones in \eqref{eq:SurfaceGauge}.

The geometries we will find are exact in $x\equiv (z\bz)^n/\rho^2$. Given this and the expansion in powers of $\kappa$, the range of $\rho$ ---the holographic radial coordinate--- needs also be small in units of the boundary curvature. This means that the expansion in $\kappa$ is also an expansion around the conformal boundary. The solutions we will write down are then analogs to Fefferman-Graham expansions, but for conically singular boundary metrics. For this reason, the field theory properties we will extract from them are approximate and belong in a UV expansion.

We will also further fix the gauge by requiring that the metric has no derivative corrections at the axis in the legs involving $dz$. That is, \emph{e.g.}, that there are no $\kappa$ corrections to eq.~\eqref{eq:gzzbaxis}. As a gauge choice, this does not affect the geometry. Its advantage is that it is straightforward to read the geometric properties of the bulk axis.

To illustrate this language, we now describe how the solution \eqref{eq:DualCone} would look like if we were finding it in this way. Starting from AdS in Poincare coordinates, and introducing a cone in the boundary by $z\rightarrow z^n$, we would write
\beq
ds^2=
\frac{d\rho^2}{\rho^2}
+ \frac{n^2(z\bz)^{n-1}dz\, d\bz}{\rho^2}
+\frac{d\zt\, d\bzt}{\rho^2}
+(n-1)ds_1^2+O\left((n-1)^2\right)\,,
\label{eq:ConeAnsatz}\eeq
with $ds_1^2$ a small correction. Taking the ansatz
\beq
ds_1^2=
f_{\rho\rho}(x)\frac{d\rho^2}{\rho^2}
+f_{z\bz}(x)\frac{n^2 (z\bz)^{n-1}dz\, d\bz}{\rho^2}
+f_{\zt\bzt}(x)\frac{d\zt\, d\bzt}{\rho^2}\,,
\label{eq:AnsatzConeCorr}\eeq
Einstein equations plus boundary conditions give:
\begin{align}
f_{\rho\rho}(x)&=-\frac{2/3}{1+x}\nonumber\\
f_{z\bz}(x)&=-\frac{2/3}{1+x}+\log\left(1+\frac{1}{x}\right)\nonumber\\
f_{\zt\bzt}(x)&=\frac{1/3}{1+x}\,,
\label{eq:ConeCorr}
\end{align}
which is indeed the $O(n-1)$ expansion of \eqref{eq:DualCone}. While the $\frac{1}{1+x}$ terms are clearly regular at $x\rightarrow 0$ and normalisable at $x\rightarrow\infty$, the $\log$ term may look problematic at the axis. However, this $\log$ is exactly what is needed to balance the zero of $(z\bz)^{n-1}$ in \eqref{eq:ConeAnsatz}, so that we are left with the finite result of eq.~\eqref{eq:gzzbaxis}. The key point is that the $(z\bz)^{n-1}$ and the $\log$ can be grouped into $(1+x)^{\frac{n-1}{n}}$ of eq.~\eqref{eq:DualCone}, exhibiting the desired contrasting behaviours at small and large $x$ ---the axis is regular in the bulk but not in the boundary.

We will find similar $\log$ divergences below, and we will have to interpret whether they reflect singular behaviour or not. They may just indicate that the ansatz following from replicating at the boundary does not capture a specific regular behaviour near the bulk axis. Sometimes this can be anticipated, like for $g_{z\bz}$ above. Consider, \emph{e.g.}, the case of $F_{\zt\bzt z\bz}\, z\,d\bz$. After replicating, this term is accompanied by a factor of $(z\bz)^{n-1}$ at the boundary. In the bulk, a minimal replica symmetric ansatz near the axis does not have such factor, and we may expect that this is the behaviour chosen by dynamics \cite{Dong:2013qoa, Camps:2013zua}. For these cases, appropriate factors of $(1+x)^{\frac{n-1}{n}}$ in the bulk ansatz can offset the appearance of logarithms in $ds_1^2$. We will anticipate so in a few cases below, except for the $R_{z \bz z\bz}$ and $Q_{\zt\bzt z\bz}$ cases; exhibiting their logarithms will be useful for understanding regularity in the more delicate case of $K_{\zt\zt z}K_{\bzt\bzt \bz}$.

Each term in the expansion in $\kappa$ of \eqref{eq:AdaptedRepl} needs its own derivative corrections. For readability, we will present all these contributions separately. It is straightforward to put them together.

We will start in \ref{sec:Kztztz} with the $O(\kappa)$ corrections, due to traceless extrinsic curvature ---we remind the reader that we exploit conformal symmetry in the boundary to drop the trace of the extrinsic curvature. We will then move on to $O(\kappa^2)$ terms. Some of these are seeded by squares of extrinsic curvatures, and others are sourced linearly by $\kappa^2$ terms in \eqref{eq:AdaptedRepl} ---including those implicit in the derivatives of the extrinsic curvature.

We will present the $O(\kappa^2)$ contributions in an order that groups them by their tensorial character in the parallel and transverse coordinates $\zt$ and $z$.  In \ref{sec:Kztztz,zt} we present the correction due to $K_{\zt\zt z,\zt}$, which might be called the $3|1$ contribution, because it has three holomorphic indices in $\zt$ and one in $z$.\footnote{Sections \ref{sec:Kztztz,zt} and \ref{sec:KztztzKztztbz} follow automatically from \ref{sec:Kztztz} and covariance in the $\zt$ coordinates, but we still present them separately for book-keeping.}  In \ref{sec:Fztbztzbz} we present the correction due to $F_{\zt\bzt z\bz}$, that may be called axial-axial because it is antisymmetric in both pairs of indices. In \ref{sec:Qztztzz} we move on to the $2|2$ term, $Q_{\zt\zt zz}$; followed in \ref{sec:Qztbztzz} and \ref{sec:KztztzKbztbztz} by the $0|2$ ones, $Q_{\zt\bzt zz}$ and $K_{\zt\zt z}K_{\bzt\bzt z}$; $2|0$ in \ref{sec:Qztztzbz},  $Q_{\zt\zt z\bz}$; $4|0$ in \ref{sec:KztztzKztztbz}, $K_{\zt\zt z}K_{\zt\zt \bz}$; and $1|1$ in \ref{sec:KPztztzbzt} and \ref{sec:Rztzzbzt}, $K_{\zt\zt z\bzt}$ and $R_{\zt z z\bz}$.  The last four contributions \ref{sec:Rztbztztbzt}-\ref{sec:KztztzKbztbztbz} are the $0|0$ ones: ${\cal R}_{\zt\bzt\zt\bzt}$, $R_{z\bz z\bz}$, $Q_{\zt\bzt z\bz}$ and $K_{\zt\zt z}K_{\bzt\bzt \bz}$. The last one has the subtlest $\log$ structure, and its regularity has consequences for the splitting problem.

Secs.~\ref{sec:Renyi} and \ref{sec:Splitting} only use the results of \ref{sec:Kztztz} and \ref{sec:Rztbztztbzt}-\ref{sec:KztztzKbztbztbz}, and some readers may want to focus on these.

Had we not used conformal symmetry to drop the trace of the extrinsic curvature, there would be five more cases at second order: $K_{\zt\bzt z,\zt}$, $K_{\zt\bzt z} K_{\zt\bzt z}$, $K_{\zt\bzt z} K_{\zt\bzt \bz}$, $K_{\zt\bzt z} K_{\zt\zt z}$, and $K_{\zt\bzt z} K_{\zt\zt \bz}$.

Reality implies that any geometry containing, \emph{e.g.}, the $0|2$ correction $K_{\zt\zt z}K_{\bzt\bzt z}$ must also contain the corrections leading to the $0|0$ correction $K_{\zt\zt z}K_{\bzt\bzt \bz}$. We will leave the latter implicit when displaying the results for the former, which means that formally we will be writing down complex metrics. This is just an artefact of the presentation in terms of $\zt$ and $z$ tensor behaviour.

For all cases we will present the expectation value of the stress tensor due to the particular squashing of the cone. This is obtained by conventional holographic methods (see sec.~\ref{sec:VacPol} for more details, discussion, and a comprehensive expression for $\langle T\rangle$).

\subsection{$K_{\zt\zt z}$}\label{sec:Kztztz}
Before replicating, the boundary metric for this term is:
\beq
ds^2_{\partial}=d\zt\, d\bzt+2\kappa\, K_{\zt\zt z}z\, d\zt^2+dz\, d\bz+O(\kappa^2)\,.
\eeq
We replicate by $z\rightarrow z^n$, leading to the bulk ansatz:
\begin{align}
ds^2=&\left(1-\frac{2}{3}\frac{(n-1)}{1+x}\right)\frac{d\rho^2}{\rho^2}\nonumber\\
&+ \left(1-\frac{2}{3}\frac{(n-1)}{1+x}\right)\frac{n^2\left({1+x}\right)^{\frac{n-1}{n}}dz\, d\bz}{\rho^{2/n}}
\qquad\nonumber\\
&+\left(1+\frac{1}{3}\frac{(n-1)}{1+x}\right)\frac{d\zt\, d\bzt+2\kappa\, K_{\zt\zt z}z^n d\zt^2}{\rho^2}\nonumber\\
&+ \kappa(n-1)ds^2_1+O\left(\kappa^2\right)+O\left((n-1)^2\right)\,,
\label{eq:metricKztztz}
\end{align}
with
\beq
ds_1^2= K_{\zt\zt z}\,f^{K2|1}_{\zt\zt}(x)\frac{2z^nd\zt^2}{\rho^2}\,.
\eeq
In \eqref{eq:metricKztztz} we choose $K_{\zt\zt z}$ to multiply the same $\frac{1}{1+x}$ factor as $g_{\zt\bzt}$ for later convenience, but that is not significant, as these factors are precisely what $f^{K2|1}_{\zt\zt}(x)$ is designed to discover. Notice that, since $ds_1^2$ is linear in $\kappa$, it has to be proportional to $K_{\zt\zt z}$; and, since this is traceless, it can only seed a $d\zt^2$ leg if we want to avoid explicit appearances of $\zt$ ---which we do because of covariance in the entangling surface. The important feature of this ansatz \eqref{eq:metricKztztz} is that the boundary metric, at $\rho\rightarrow 0$, is different from the one in \eqref{eq:DualCone} by the factor of $K_{\zt\zt z} z^n$ of \eqref{eq:AdaptedRepl}. 

Einstein's equations lead an ODE for $f^{K2|1}_{\zt\zt}(x)$, whose normalisable solution is:
\beq
f^{K2|1}_{\zt\zt}(x)=C_{K2|1}\left(\frac{1}{x}-\log\left(1+\frac{1}{x}\right)\right)\,,
\eeq
with $C_{K2|1}$ an integration constant. $1/x$ is singular at the axis, and regularity sets $C_{K2|1}=0$.

Restoring the AdS radius $\ell$, the stress tensor reads\footnote{The $O(\kappa^0)$ term agrees with \cite{Smolkin:2014hba}. The $O(\kappa)$ term agrees with \cite{Dong:2016wcf}.}
\begin{align}
\langle T\rangle=\frac{\ell^3(n-1)}{4\pi G(z\bz)^{2n}}\frac{-1}{6}\left(d\zt\,d\bzt+\kappa\, K_{\zt\zt z}z^nd\zt^2+(z\bz)^{n-1}\left(-dz\,d\bz+\frac{\bz^2dz^2+z^2d\bz^2}{z\bz}\right)\right)
\quad&\nonumber\\
+O(\kappa^2)+O\left((n-1)^2\right)&\,.
\label{eq:TKbzbzz}
\end{align}
\subsection{$K_{\zt\zt z,\zt}$}\label{sec:Kztztz,zt}
This term follows directly the one we just analysed, from covariance in the $\zt$ coordinate. Its boundary metric is, before replicating
\beq
ds^2_{\partial}=d\zt\, d\bzt+2\kappa^2 K_{\zt\zt z,\zt}\,\zt\, z\, d\zt^2+dz\, d\bz+O(\kappa^2)\,.
\eeq
The bulk dual is
\begin{align}
ds^2=&\left(1-\frac{2}{3}\frac{(n-1)}{1+x}\right)\frac{d\rho^2}{\rho^2}\nonumber\\
&+ \left(1-\frac{2}{3}\frac{(n-1)}{1+x}\right)\frac{n^2\left({1+x}\right)^{\frac{n-1}{n}}dz\, d\bz}{\rho^{2/n}}
\qquad\nonumber\\
&+\left(1+\frac{1}{3}\frac{(n-1)}{1+x}\right)\frac{d\zt\, d\bzt+2\kappa^2 K_{\zt\zt z,\zt}\,\zt\,z^n d\zt^2}{\rho^2}\nonumber\\
&+ \kappa^2(n-1)ds^2_1+O\left(\kappa^2\right)+O\left((n-1)^2\right)\,,
\label{eq:metricKztztz,zt}
\end{align}
with
\beq
ds_1^2= K_{\zt\zt z,\zt}\,\zt\,f^{K3|1}_{\zt\zt}(x)\frac{2z^nd\zt^2}{\rho^2}\,.
\eeq

Einstein's equations lead the same ODE for $f^{K3|1}_{\zt\zt}(x)$ as for $f^{K2|1}_{\zt\zt}(x)$, and we pick the  regular solution: $f^{K3|1}_{\zt\zt}(x)=0$

The contribution to the stress tensor reads
\beq
\langle T\rangle=\frac{\ell^3(n-1)}{4\pi G(z\bz)^{2n}}\frac{-1}{6}\left(\kappa^2\, K_{\zt\zt z,\zt}\,\zt\,z^nd\zt^2\right)\,,
\eeq
and follows from \eqref{eq:TKbzbzz} by $K_{\zt\zt z}\rightarrow K_{\zt\zt z,\zt}\,\zt$, as dictated by covariance in $\zt$.

\subsection{$F_{\zt\bzt z\bz}$}\label{sec:Fztbztzbz}
The boundary metric for this term is, before replicating,
\beq
ds^2_{\partial}=d\zt\, d\bzt+dz\, d\bz-\kappa^2F_{\zt\bzt z\bz} (z\,d\bz-\bz\, dz)(\zt\,d\bzt-\bzt\, d\zt)+O(\kappa^3)\,.
\eeq
After $z\rightarrow z^n$, a natural ansatz for the bulk is:
\begin{align}
ds^2=&\left(1-\frac{2}{3}\frac{(n-1)}{1+x}\right)\frac{d\rho^2}{\rho^2}\nonumber\\
&+ \left(1-\frac{2}{3}\frac{(n-1)}{1+x}\right)\frac{n^2\left({1+x}\right)^{\frac{n-1}{n}}dz\, d\bz-n\,\left({1+x}\right)^{\frac{n-1}{n}}\kappa^2F_{\zt\bzt z\bz} (z\,d\bz-\bz dz)(\zt\,d\bzt-\bzt d\zt)}{\rho^{2/n}}
\qquad\nonumber\\
&+\left(1+\frac{1}{3}\frac{(n-1)}{1+x}\right)\frac{d\zt\, d\bzt}{\rho^2}\nonumber\\
&+ \kappa^2(n-1)ds^2_1+O(\kappa^3)+O\left((n-1)^2\right)\,,
\end{align}
with
\beq
ds_1^2=F_{\zt\bzt z\bz}\, f^{F\textrm{aa}}_{z\zt}(x)\frac{(z\,d\bz-\bz\, dz)(\zt\,d\bzt-\bzt\, d\zt)}{\rho^{2/n}}\,.
\eeq

Notice the factor of $(1+x)^{\frac{n-1}{n}}$, anticipating different behaviours of the $F_{\zt\bzt z\bz}$ term in the boundary and bulk axes. In the boundary, we require the $(z\bz)^{n-1}$ behaviour from the replica trick \eqref{eq:AdaptedRepl}; in the bulk, a minimal guess suggests $(z\bz)^0$.

The normalisable solution we find is:
\beq
f^{F\textrm{aa}}_{z\zt}(x)=C_{F\textrm{aa}}\left(\frac{1}{x}-\log\left(1+\frac{1}{x}\right)\right)\,,
\eeq
and it should be set to zero because it is not regular at the bulk axis $x\rightarrow 0$.

The $O(\kappa^2)$ contribution to the stress tensor coming from this term is:
\beq
\langle T\rangle=-\frac{\ell^3(n-1)}{4\pi G(z\bz)^{2n}}\frac{\kappa^2F_{\zt\bzt z\bz}}{2}
(z\bz)^{n-1}(z\,d\bz-\bz\, dz)(\zt\,d\bzt-\bzt\, d\zt)\,.
\eeq
\subsection{$Q_{\zt\zt zz}$}\label{sec:Qztztzz}
This one comes from
\beq
ds^2_{\partial}=d\zt\, d\bzt+dz\, d\bz+\kappa^2Q_{\zt\zt zz}z^2 d\zt^2+O(\kappa^3)\,.
\eeq
Then,
\begin{align}
ds^2=&\left(1-\frac{2}{3}\frac{(n-1)}{1+x}\right)\frac{d\rho^2}{\rho^2}\nonumber\\
&+ \left(1-\frac{2}{3}\frac{(n-1)}{1+x}\right)\frac{n^2\left({1+x}\right)^{\frac{n-1}{n}}dz\, d\bz}{\rho^{2/n}}
\qquad\nonumber\\
&+\left(1+\frac{1}{3}\frac{(n-1)}{1+x}\right)\frac{d\zt\, d\bzt+\kappa^2 Q_{\zt\zt zz}z^{2n}d\zt^2}{\rho^2}
\nonumber\\
&+ \kappa^2(n-1)ds^2_1+O(\kappa^3)+O\left((n-1)^2\right)\,,
\end{align}
with
\beq
ds_1^2=Q_{\zt\zt zz} f^{Q2|2}_{\zt\zt}(x)\frac{z^{2n}}{\rho^2}d\zt^2\,.
\eeq

The normalisable solution to Einstein equations is:
\beq
f^{Q2|2}_{\zt\zt}(x)=\frac{C_{Q2|2}}{x^2}\,,
\eeq
which regularity sets to zero.

The contribution to the stress tensor vanishes:
\beq
\langle T\rangle=0\,.
\eeq
\subsection{$Q_{\zt\bzt zz}$}\label{sec:Qztbztzz}
This one comes from
\beq
ds^2_{\partial}=(1+2\kappa^2 Q_{\zt\bzt zz} z^2)d\zt\, d\bzt+dz\, d\bz+O(\kappa^3)\,.
\label{eq:BoundaryQztbztzz}\eeq
Then,
\begin{align}
ds^2=&\left(1-\frac{2}{3}\frac{(n-1)}{1+x}\right)\frac{d\rho^2}{\rho^2}\nonumber\\
&+ \left(1-\frac{2}{3}\frac{(n-1)}{1+x}\right)\frac{n^2\left({1+x}\right)^{\frac{n-1}{n}} dz\, d\bz}{\rho^{2/n}}
\nonumber\\
&-\left(1-\frac{2}{3}\frac{(n-1)}{1+x}\right)4\kappa^2Q_{\zt\bzt zz}\frac{z^n\, n\, z^{n-1}dz\, d\rho}{\rho}
\nonumber\\
&+\left(1+\frac{1}{3}\frac{(n-1)}{1+x}\right)\frac{\left(1+2\kappa^2Q_{\zt\bzt zz} z^{2n}\right)d\zt\, d\bzt}{\rho^2}
\nonumber\\
&+ \kappa^2(n-1)ds^2_1+O(\kappa^3)+O\left((n-1)^2\right)\,,
\label{eq:metricQztbztzz}
\end{align}
with
\beq
ds_1^2=2Q_{\zt\bzt zz} \frac{z^{2n}}{\rho^2}\left(f^{Q0|2}_{\rho\rho}(x)d\rho^2+f^{Q0|2}_{z\bz}(x)(z\bz)^{n-1}dz\,d\bz+f^{Q0|2}_{\zt\bzt}(x)d\zt\, d\bzt+f^{Q0|2}_{z\rho}(x)\frac{\rho}{z}dz\, d\rho\right)\,.
\eeq

This case has new features compared to the previous two.

First, \eqref{eq:metricQztbztzz} has a $\kappa^2$ term in the $d\rho\, dz$ leg that remains finite as $n\rightarrow 1$. This term does not change the boundary metric and therefore is a bulk response to $Q_{\zt\bzt zz}$, even before introducing any conical singularity. Its origin is, in fact, well known. In the Fefferman-Graham expansion it is the Shouten term,\footnote{
The Fefferman-Graham expansion of AAdS spacetimes reads, for small $\rho$ \cite{deHaro:2000vlm}:
$$
ds^2=\frac{d\rho^2}{\rho^2}+\frac{g_{\mu\nu}(\rho,x) dx^\mu dx^\nu}{\rho^2}\,,
\qquad\qquad
g_{\mu\nu}(\rho,x)={}^{(0)}g_{\mu\nu}(x)+{}^{(2)}g_{\mu\nu}(x)\rho^2+\cdots\,.
$$
${}^{(2)}g_{\mu\nu}=\frac{1}{d-2}\left({}^{(0)}R_{\mu\nu}-\frac{{}^{(0)}R}{2(d-1)}{}^{(0)}g_{\mu\nu}\right)$ is the Shouten tensor  of the boundary metric (for boundary dimension $d>2$). Notice that the Fefferman-Graham expansion is a derivative expansion, so the expansion in $\kappa$ has a FG character.\label{footnote:FG}}
 which is indeed non-zero for \eqref{eq:BoundaryQztbztzz}. As explained in the opening of this section, our gauge demands that $g_{zz}$ vanishes on the bulk axis. This places this Shouten term in the $d\rho\, dz$ leg (in Fefferman-Graham coordinates this would have had a $dz^2$ leg).

Second, the scalar character of $Q_{\zt\bzt zz}$ in the $\zt$ directions allows for many more terms in $ds_1^2$ compared to earlier. In fact, covariance would also allow for a $dz^2$ leg that we have not written down. As it turns out, a gauge transformation can move this correction from the $dz^2$ leg to $dz\, d\rho$, and, as explained, our gauge fixing places it in the latter.

There are two normalizable zero modes to this ansatz:
\begin{align}
f^{Q0|2}_{\rho\rho}(x)&=-2C^{0|2}_1\left(\frac{1}{x}+\frac{1}{x^2}\right)\nonumber\\
f^{Q0|2}_{\zt\bzt}(x)&=C^{0|2}_1\left(\frac{1}{x}+\frac{1}{x^2}\right)\nonumber\\
f^{Q0|2}_{z\bz}(x)&=-\frac{C^{0|2}_2}{4}\frac{1}{x^2}+C^{0|2}_1\left(\frac{1}{x}-\frac{1}{2}\frac{1}{x^2}\right)\nonumber\\
f^{Q0|2}_{z\rho}(x)&=C^{0|2}_2\frac{1}{x}\,,
\label{eq:zeroModeQztbztzz}\end{align}
and they are banished by regularity at the axis.

The regular solution we find is:
\begin{align}
f^{Q0|2}_{\rho\rho}(x)&=\frac{4}{9}\left(\frac{1}{1+x}-\frac{1}{(1+x)^2}\right)\nonumber\\
f^{Q0|2}_{\zt\bzt}(x)&=\frac{2}{9}\left(-\frac{1}{1+x}+\frac{1}{(1+x)^2}\right)\nonumber\\
f^{Q0|2}_{z\bz}(x)&=\frac{-2}{9}\left(\frac{1}{1+x}+\frac{2}{(1+x)^2}\right)\nonumber\\
f^{Q0|2}_{z\rho}(x)&=0\,.
\end{align}

The contribution of $Q_{\zt\bzt zz}$ to the stress tensor is:
\beq
\langle T\rangle=\frac{\ell^3(n-1)}{4\pi G(z\bz)^{2n}}\frac{\kappa^2Q_{\zt\bzt zz}\,z^{2n}}{9}
\left(d\zt\, d\bzt+(z\bz)^{n-1}\left(-4dz\,d\bz+2\frac{z^2 d\bz^2+3\bz^2 dz^2}{z\bz}\right)\right)\,.
\eeq

\subsection{$K_{\zt\zt z}K_{\bzt\bzt z}$}\label{sec:KztztzKbztbztz}
This one comes from
\beq
ds^2_{\partial}=d\zt\, d\bzt+2\kappa\left(K_{\zt\zt z}z\,d\zt^2+K_{\bzt\bzt z}z\,d\bzt^2\right)+dz\, d\bz+O(\kappa^3)\,.
\eeq
Then,
\begin{align}
ds^2=&\left(1-\frac{2}{3}\frac{(n-1)}{1+x}\right)\frac{d\rho^2}{\rho^2}\nonumber\\
&+ \left(1-\frac{2}{3}\frac{(n-1)}{1+x}\right)\frac{\left({1+x}\right)^{\frac{n-1}{n}}n^2dz\, d\bz}{\rho^{2/n}}
\nonumber\\
&+\left(1-\frac{2}{3}\frac{(n-1)}{1+x}\right)8\kappa^2K_{\zt\zt z}K_{\bzt\bzt z}\frac{z^n\,n\,z^{n-1} dz\, d\rho}{\rho}
\nonumber\\
&+\left(1+\frac{1}{3}\frac{(n-1)}{1+x}\right)\frac{d\zt\, d\bzt+2\kappa
\left(K_{\zt\zt z}z^n d\zt^2+K_{\bzt\bzt z}z^n d\bzt^2\right)}{\rho^2}
\nonumber\\
&+ \kappa^2(n-1)ds^2_1+O(\kappa^3)+O\left((n-1)^2\right)\,,
\label{eq:metricKztztzKbztbztz}
\end{align}
with
\beq
ds_1^2=2K_{\zt\zt z}K_{\bzt\bzt z} \frac{z^{2n}}{\rho^2}\left(f^{K0|2}_{\rho\rho}(x)d\rho^2+f^{K0|2}_{z\bz}(x)(z\bz)^{n-1}dz\,d\bz+f^{K0|2}_{\zt\bzt}(x)d\zt\, d\bzt+f^{K0|2}_{z\rho}(x)\frac{\rho}{z}dz\, d\rho\right)\,.
\eeq
Again, the third line in \eqref{eq:metricKztztzKbztbztz} comes from the Shouten tensor, and there is no $dz^2$ leg because of gauge fixing.

Discarding the same zero mode as earlier, \eqref{eq:zeroModeQztbztzz}, the regular solution we find is:
\begin{align}
f^{K0|2}_{\rho\rho}(x)=&\frac{2}{9}\left(\frac{-7}{1+x}+\frac{4}{(1+x)^2}\right)\nonumber\\
f^{K0|2}_{\zt\bzt}(x)=&\frac{1}{9}\left(\frac{7}{1+x}-\frac{4}{(1+x)^2}\right)\nonumber\\
f^{K0|2}_{z\bz}(x)=&\frac{1}{9x}\left(7+\frac{10}{1+x}-\frac{8}{(1+x)^2}\right)-\frac{1}{x^2}\log(1+x)\nonumber\\
f^{K0|2}_{z\rho}(x)=&\frac{4}{x}\log(1+x)\,.
\end{align}

The contribution to the stress tensor:
\beq
\langle T\rangle=\frac{\ell^3(n-1)}{4\pi G(z\bz)^{2n}}\frac{\kappa^2K_{\zt\zt z}K_{\bzt\bzt z} z^{2n}}{9}
\left(-5d\zt\, d\bzt+(z\bz)^{n-1}\left(17dz\,d\bz-\frac{7z^2 d\bz^2+30\bz^2 dz^2}{z\bz}\right)\right)\,.
\eeq
\subsection{$Q_{\zt\zt z\bz}$}\label{sec:Qztztzbz}
This one comes from
\beq
ds^2_{\partial}=d\zt\, d\bzt+2\kappa^2Q_{\zt\zt z\bz} z \bz\, d\zt^2+dz\, d\bz+O(\kappa^3)\,.
\eeq
Then,
\begin{align}
ds^2=&\left(1-\frac{2}{3}\frac{(n-1)}{1+x}\right)\frac{d\rho^2}{\rho^2}\nonumber\\
&+ \left(1-\frac{2}{3}\frac{(n-1)}{1+x}\right)\left[\frac{n^2\left({1+x}\right)^{\frac{n-1}{n}}dz\, d\bz+2\kappa^2Q_{\zt\zt z\bz}\left({1+x}\right)^{\frac{n-1}{n}}z\bz\,d\zt^2}{\rho^{2/n}}+2\kappa^2 Q_{\zt\zt z\bz}d\zt^2\right]
\qquad\nonumber\\
&+\left(1+\frac{1}{3}\frac{(n-1)}{1+x}\right)\frac{d\zt\, d\bzt}{\rho^2}
\nonumber\\
&+ \kappa^2(n-1)ds^2_1+O\left(\kappa^3\right)+O\left((n-1)^2\right)\,,
\end{align}
with
\beq
ds_1^2=2Q_{\zt\zt z\bz} f^{Q2|0}_{\zt\zt}(x) d\zt^2\,.
\eeq
Again, notice a Shouten correction in the brackets. Notice also that we have anticipated a change of behaviour of $Q_{\zt\zt z\bz}$ from the boundary, $(z\bz)^{n}$, to the bulk axis, $z\bz$.

$f^{Q2|0}_{\zt\zt}(x)=0$ is the regular solution we seek. There is also one singular normalisable zero mode:
\beq
f^{Q2|0}_{\zt\zt}(x)=C^{Q2|0}\left(1-(1+x)\log\left(1+\frac{1}{x}\right)\right)\,.
\eeq

Stress tensor contribution:
\beq
\langle T\rangle=\frac{\ell^3(n-1)}{4\pi G(z\bz)^{2n}}\frac{-\kappa^2Q_{\zt\zt z\bz} (z \bz)^n}{3}
d\zt^2\,.
\eeq

\subsection{$K_{\zt\zt z}K_{\zt\zt \bz}$}\label{sec:KztztzKztztbz}
This one comes from
\beq
ds^2_{\partial}=d\zt\, d\bzt+2\kappa\left(K_{\zt\zt z}z+K_{\zt\zt \bz}\bz\right) d\zt^2+dz\, d\bz+O(\kappa^3)\,.
\eeq
Then,
\begin{align}
ds^2=&\left(1-\frac{2}{3}\frac{(n-1)}{1+x}\right)\frac{d\rho^2}{\rho^2}\nonumber\\
&+ \left(1-\frac{2}{3}\frac{(n-1)}{1+x}\right)\frac{n^2\left({1+x}\right)^{\frac{n-1}{n}}dz\, d\bz}{\rho^{2/n}}
\qquad\nonumber\\
&+\left(1+\frac{1}{3}\frac{(n-1)}{1+x}\right)\frac{d\zt\, d\bzt+2\kappa \left(K_{\zt\zt z}z^n+K_{\zt\zt \bz}\bz^n\right)\,d\zt^2}{\rho^2}
\nonumber\\
&+ \kappa^2(n-1)ds^2_1+O\left(\kappa^3\right)+O\left((n-1)^2\right)\,,
\end{align}
but $ds_1^2=0$ because the four legs in $\zt$ of $K_{\zt\zt z}K_{\zt\zt \bz}$ would force at least two contractions with $\zt$, which clashes with covariance in $\zt$. This case, as the one in \ref{sec:Kztztz,zt}, follows from the one in \ref{sec:Kztztz}.

Therefore,
\beq
\langle T\rangle=0\,.
\eeq
\subsection{$K_{\zt\zt z,\bzt}$}\label{sec:KPztztzbzt}
This one comes from
\beq
ds^2_{\partial}=d\zt\, d\bzt+2\kappa^2 K_{\zt\zt z,\bzt}\,\bzt\,z \, d\zt^2+dz\, d\bz+O(\kappa^3)\,.
\eeq
Then,
\begin{align}
ds^2=&\left(1-\frac{2}{3}\frac{(n-1)}{1+x}\right)\frac{d\rho^2}{\rho^2}\nonumber\\
&+ \left(1-\frac{2}{3}\frac{(n-1)}{1+x}\right)\frac{n^2\left({1+x}\right)^{\frac{n-1}{n}}dz\, d\bz}{\rho^{2/n}}
\qquad\nonumber\\
&+\left(1+\frac{1}{3}\frac{(n-1)}{1+x}\right)\frac{d\zt\, d\bzt+2\kappa^2 K_{\zt\zt z,\bzt}\,\bzt\,z^n\,d\zt^2}{\rho^2}
\nonumber\\
&+4\kappa^2 K_{\zt\zt z,\bzt}\frac{z^n\,d\zt\,d\rho}{\rho}\nonumber\\
&+ \kappa^2(n-1)ds^2_1+O\left(\kappa^3\right)+O\left((n-1)^2\right)\,,
\end{align}
with
\beq
ds_1^2=2K_{\zt\zt z,\bzt}\left(f^{K1|1}_{\zt z}(x)\,z^{n-1}dz\,d\zt
+f^{K1|1}_{\zt\rho}(x) \frac{z^n\,d\zt\,d\rho}{\rho}
\right)\,.
\eeq
Notice a Shouten correction in the fourth line.

There are two normalisable zero-modes:
\begin{align}
f^{K1|1}_{\zt z}(x)&=C^{1|1}_2\nonumber\\
f^{K1|1}_{\zt\rho}(x)&=2C^{1|1}_2+\frac{C^{1|1}_1}{x}\,.
\end{align}
We set $C^{1|1}_1=0$ for regularity at the axis, and $C^{1|1}_2=0$ with the gauge condition $g_{z\mu}=0$ at the axis.

The regular particular solution is:
\begin{align}
f^{K1|1}_{\zt z}(x)&=-\frac{1}{2}\frac{x}{1+x}\\
f^{K1|1}_{\zt\rho}(x)&=-1+\frac{1}{6}\frac{1}{1+x}\,.
\end{align}

Stress tensor contribution:
\beq
\langle T\rangle=\frac{\ell^3(n-1)}{4\pi G(z\bz)^{2n}}\frac{-\kappa^2 K_{\zt\zt z,\bzt}\, z^n}{6}
\left(\bzt\, d\zt^2+(z\bz)^{n-1}\left(\frac{3}{2} z\, d\bz\,d\zt-2\bz\, dz\,d\zt\right)\right)\,.
\eeq
The first of these terms follows from covariance in \eqref{eq:TKbzbzz}. The rest are new.

\subsection{$R_{\zt z z\bz}$}\label{sec:Rztzzbzt}
This one comes from
\beq
ds^2_{\partial}=d\zt\, d\bzt+dz\, d\bz-\frac{4}{3}\kappa^2 R_{\zt z z\bz}\,z(\bz\,dz-z\, d\bz)d\zt+O(\kappa^3)\,.
\eeq
Then,
\begin{align}
ds^2=&\left(1-\frac{2}{3}\frac{(n-1)}{1+x}\right)\frac{d\rho^2}{\rho^2}\nonumber\\
&+ \left(1-\frac{2}{3}\frac{(n-1)}{1+x}\right)\frac{n^2\left({1+x}\right)^{\frac{n-1}{n}}dz\, d\bz}{\rho^{2/n}}
\qquad\nonumber\\
&+\left(1+\frac{1}{3}\frac{(n-1)}{1+x}\right)\frac{d\zt\, d\bzt}{\rho^2}
\nonumber\\
&-\left(1-\frac{2}{3}\frac{(n-1)}{1+x}\right)\frac{4}{3}\kappa^2 R_{\zt z z\bz}\frac{z^n\,n^2\,(1+x)^{\frac{n-1}{n}}(\bz\,dz-z\, d\bz)d\zt}{\rho^{2/n}}
\nonumber\\
&+4\kappa^2 R_{\zt z z\bz}\frac{z^{n}d\rho\,d\zt}{\rho}
\nonumber\\
&+ \kappa^2(n-1)ds^2_1+O\left(\kappa^3\right)+O\left((n-1)^2\right)\,,
\end{align}
with
\beq
ds_1^2=R_{\zt z z\bz}\left(f^{R1|1a}_{\zt z}(x)\frac{z^n(\bz\,dz-z\,d\bz)\,d\zt}{\rho^{2/n}}
+f^{R1|1a}_{\zt\rho}(x) \frac{z^n d\zt\,d\rho}{\rho}
\right)\,.
\eeq
Notice a Shouten correction in the fourth line.

There are two normalisable zero-modes:
\begin{align}
f^{R1|1a}_{\zt z}(x)&=\frac{1}{6}\frac{C^{1|1a}_1}{x^{3/2}}\nonumber\\
f^{R1|1a}_{\zt \rho}(x)&=\frac{C^{1|1a}_1}{x^{1/2}}+\frac{C^{1|1a}_2}{x}\,.
\end{align}
We set $C^{1|1a}_1=0$ for regularity at the axis, and $C^{1|1a}_2$ with the gauge condition $g_{z\mu}=0$.

The regular particular solution is:
\begin{align}
f^{R1|1a}_{\zt z}(x)&=\frac{5}{6x}\left(\frac{1}{1+x}-\frac{\arctan\sqrt{x}}{\sqrt{x}}\right)\\
f^{R1|1a}_{\zt \rho}(x)&=\frac{4}{3}\frac{1}{1+x}-5\frac{\arctan\sqrt{x}}{\sqrt{x}}\,.
\end{align}

Stress tensor contribution:
\beq
\langle T\rangle=\frac{\ell^3(n-1)}{4\pi G(z\bz)^{2n}}\frac{\kappa^2 R_{\zt z z\bz}\, z^n}{36}
(z\bz)^{n-1}\left(8\bz\, dz\,d\zt+13 z\, d\bz\,d\zt\right)\,.
\eeq

\subsection{${\cal R}_{\zt\bzt\zt\bzt}$}\label{sec:Rztbztztbzt}
This one comes from
\beq
ds^2_{\partial}=d\zt\, d\bzt-\frac{1}{3}\kappa^2{\cal R}_{\zt\bzt\zt\bzt} \left(\bzt\, d\zt-\zt\, d\bzt\right)^2+dz\, d\bz+O(\kappa^3)\,.
\eeq
Then,
\begin{align}
ds^2=&\left(1-\frac{2}{3}\frac{(n-1)}{1+x}\right)\left[1-\frac{4}{3}\kappa^2 \rho^2\right]\frac{d\rho^2}{\rho^2}\nonumber\\
&+ \left(1-\frac{2}{3}\frac{(n-1)}{1+x}\right)\frac{n^2\left({1+x}\right)^{\frac{n-1}{n}}dz\, d\bz}{\rho^{2/n}}
\qquad\nonumber\\
&+\left(1+\frac{1}{3}\frac{(n-1)}{1+x}\right)\left[\frac{d\zt\, d\bzt-\frac{1}{3}\kappa^2{\cal R}_{\zt\bzt\zt\bzt} \left(\bzt\, d\zt-\zt\, d\bzt\right)^2}{\rho^2}+2\kappa^2{\cal R}_{\zt\bzt\zt\bzt}\,d\zt\, d\bzt\right]
\nonumber\\
&+ \kappa^2(n-1)ds^2_1+O\left((\kappa\,x)^3\right)+O\left((n-1)^2\right)\,,
\end{align}
with
\beq
ds_1^2=-2{\cal R}_{\zt\bzt\zt\bzt}\left(f^{R\zt}_{\rho\rho}(x)d\rho^2+f^{R\zt}_{z\bz}(x)(z\bz)^{n-1}dz\,d\bz+f^{R\zt}_{\zt\bzt}(x)d\zt\, d\bzt\right)\,.
\eeq
Notice again Shouten corrections within the brackets (see footnote \ref{footnote:FG}).

There are two normalizable zero modes to this ansatz, that we will omit henceforth ---also for the three other $0|0$ cases that follow. They look
\begin{align}
f^{0|0}_{\rho\rho}(x)=&2C^{0|0}_2-2C^{0|0}_1\left(\log x-1\right)\nonumber\\
f^{0|0}_{\zt\bzt}(x)=&-C^{0|0}_2+C^{0|0}_1\log x&\nonumber\\
f^{0|0}_{z\bz}(x)=&-C^{0|0}_2+C^{0|0}_1\log x\,.
\label{eq:SSZeroModes}\end{align}
$C^{0|0}_2=0$ with the gauge condition of no corrections in $\kappa$ to $g_{z\bz}$ at the axis, and $C^{0|0}_1=0$ with regularity. Notice that $C^{0|0}_1\neq 0$ would result in a logarithmic divergence at the axis in, \emph{e.g.}, the $dz\, d\bz$ leg. This leg has no $\kappa^2$ term surviving the $n\rightarrow 1$ limit, so no candidate to absorb the logarithmic divergence in a change of behaviour of the exponent, as is the case for the logarithmic divergences that we have been allowing.

The regular solution is:
\begin{align}
f^{R\zt}_{\rho\rho}(x)=&\frac{2}{27}\left(10-\frac{5}{1+x}+\frac{2}{(1+x)^2}-10\log(1+x)\right)\nonumber\\
f^{R\zt}_{\zt\bzt}(x)=&\frac{-1}{27}\left(-\frac{5}{1+x}+\frac{2}{(1+x)^2}-10\log(1+x)\right)&\nonumber\\
f^{R\zt}_{z\bz}(x)=&\frac{1}{27}\left(-\frac{4}{1+x}+\frac{4}{(1+x)^2}+10\log(1+x)\right)\,.
\label{eq:solRztbztztbzt}\end{align}

Stress tensor contribution:
\beq
\langle T\rangle=\frac{\ell^3(n-1)}{4\pi G(z\bz)^{2n}}\frac{\kappa^2{\cal R}_{\zt\bzt \zt\bzt}}{18}
\left((\zt\, d\bzt-\bzt\, d\zt)^2-\frac{10}{3}(z\bz)^{2(n-1)}(z^2 d\bz^2+\bz^2 dz^2)\right)\,.
\eeq

\subsection{$R_{z\bz z\bz}$}\label{sec:Rzbzzbz}
This one comes from
\beq
ds^2_{\partial}=d\zt\, d\bzt+(1+2\kappa^2R_{z\bz z\bz}\,z\bz)dz\, d\bz+O(\kappa^3)\,.
\eeq
Then, after replicating, we write the following ansatz
\begin{align}
ds^2=&\left(1-\frac{2}{3}\frac{(n-1)}{1+x}\right)\left[1+\frac{8}{3}\kappa^2 R_{z\bz z\bz}\rho^2\right]\frac{d\rho^2}{\rho^2}\nonumber\\
&+ \left(1-\frac{2}{3}\frac{(n-1)}{1+x}\right)\frac{n^2\left({1+x}\right)^{\frac{n-1}{n}}\left(1+2\kappa^2 R_{z\bz z\bz}(z\bz)^n\right)dz\, d\bz}{\rho^{2/n}}
\qquad\nonumber\\
&+\left(1+\frac{1}{3}\frac{(n-1)}{1+x}\right)\left[\frac{d\zt\, d\bzt}{\rho^2}-2\kappa^2  R_{z\bz z\bz}\,d\zt\, d\bzt\right]
\nonumber\\
&+ \kappa^2(n-1)ds^2_1+O(\kappa^3)+O\left((n-1)^2\right)\,,
\label{eq:metricRzzbzzb}
\end{align}
with
\beq
ds_1^2=2R_{z\bz z\bz}\left(f^{Rz}_{\rho\rho}(x)d\rho^2+f^{Rz}_{z\bz}(x)(z\bz)^{n-1}dz\,d\bz+f^{Rz}_{\zt\bzt}(x)d\zt\, d\bzt\right)\,.
\eeq
As earlier, Shouten corrections appear in brackets. Notice that we have not written down $(1+x)^{\frac{n-1}{n}}$ factor in the $R_{z\bz z\bz}$ factor second line, even though we anticipate it. The reason for not writing it will be clear in sec.~\ref{sec:KztztzKbztbztbz}.

The regular solution we find is:
\begin{align}
f^{Rz}_{\rho\rho}(x)=&\frac{2}{27}\left(-1-\frac{13}{1+x}+\frac{7}{(1+x)^2}-17\log(1+x)\right)\nonumber\\
f^{Rz}_{\zt\bzt}(x)=&\frac{-1}{27}\left(-18-\frac{13}{1+x}+\frac{7}{(1+x)^2}-17\log(1+x)\right)&\nonumber\\
f^{Rz}_{z\bz}(x)=&\frac{1}{27}\left(-9-\frac{5}{1+x}+\frac{14}{(1+x)^2}+17\log(1+x)\right)+x\log\left(1+\frac{1}{x}\right)\,.
\label{eq:solRzbzzbz}\end{align}
Notice the appearance of the logarithm, that could be absorbed in the ansatz by writing, instead of  $R_{z\bz z\bz}\, (z\bz)^n$ in the second line of \eqref{eq:metricRzzbzzb}, $R_{z\bz z\bz}\, z\bz\, (1+x)^{\frac{n-1}{n}}\rho^{2\frac{n-1}{n}}$. Thus, this logarithm does not reflect singular behaviour.

Stress tensor contribution:
\beq
\langle T\rangle=\frac{\ell^3(n-1)}{4\pi G(z\bz)^{2n}}\frac{\kappa^2R_{z \bz z \bz}(z\bz)^n}{3}
\left(2d\zt\, d\bzt+(z\bz)^{n-1}\left(-dz\, d\bz+\frac{4}{9}\frac{\bz^2 dz^2+z^2d\bz^2}{z\bz}\right)
\right)\,.
\eeq

\subsection{$Q_{\zt\bzt z\bz}$}\label{sec:Qztbztzbz}
This one comes from
\beq
ds^2_{\partial}=(1+4\kappa^2 Q_{\zt\bzt z\bz} z \bz)d\zt\, d\bzt+dz\, d\bz+O(\kappa^3)\,.
\eeq
Then, after replicating,
\begin{align}
ds^2=&\left(1-\frac{2}{3}\frac{(n-1)}{1+x}\right)\left[\frac{d\rho^2}{\rho^2}+\frac{8}{3}\kappa^2 Q_{\zt\bzt z\bz}\,d\rho^2\right]\nonumber\\
&+ \left(1-\frac{2}{3}\frac{(n-1)}{1+x}\right)\frac{\left({1+x}\right)^{\frac{n-1}{n}}n^2dz\, d\bz}{\rho^{2/n}}
\qquad\nonumber\\
&+\left(1+\frac{1}{3}\frac{(n-1)}{1+x}\right)\frac{\left(1+4\kappa^2 Q_{\zt\bzt z\bz}(z\bz)^n\right)d\zt\, d\bzt}{\rho^2}
\nonumber\\
&+ \kappa^2(n-1)ds^2_1+O(\kappa^3)+O\left((n-1)^2\right)\,,
\label{eq:metricQztbztzbz}
\end{align}
with
\beq
ds_1^2=4Q_{\zt\bzt z\bz} \left(f^{Q0|0}_{\rho\rho}(x)d\rho^2+f^{Q0|0}_{z\bz}(x)(z\bz)^{n-1}dz\,d\bz+f^{Q0|0}_{\zt\bzt}(x)d\zt\, d\bzt\right)\,.
\eeq
Notice again the Shouten correction inside the brackets, and the absence of a likely $(1+x)^{\frac{n-1}{n}}$ factor in the third line of \eqref{eq:metricQztbztzbz}.

After discarding the singular zero mode of \eqref{eq:SSZeroModes}, we find the regular solution:
\begin{align}
f^{Q0|0}_{\rho\rho}(x)=&\frac{2}{27}\left(10-\frac{5}{1+x}+\frac{2}{(1+x)^2}-10\log(1+x)\right)\nonumber\\
f^{Q0|0}_{\zt\bzt}(x)=&\frac{-1}{27}\left(27-\frac{5}{1+x}+\frac{2}{(1+x)^2}-10\log(1+x)\right)+x\log\left(1+\frac{1}{x}\right)&\nonumber\\
f^{Q0|0}_{z\bz}(x)=&\frac{1}{27}\left(-\frac{4}{1+x}+\frac{4}{(1+x)^2}+10\log(1+x)\right)
\label{eq:solQztbztzbz}\end{align}
The apparently singular $\log$ in $f^{Q0|0}_{\zt\bzt}(x)$ can again be absorbed in the ansatz \eqref{eq:metricQztbztzbz} by replacing $Q_{\zt\bzt z\bz}\, (z\bz)^n$ in the third line by $Q_{\zt\bzt z\bz}\, z\bz\, (1+x)^{\frac{n-1}{n}}\rho^{2\frac{n-1}{n}}$.

The contribution to the stress tensor:
\beq
\langle T\rangle=\frac{\ell^3(n-1)}{4\pi G(z\bz)^{2n}}\frac{-2\kappa^2 Q_{\zt\bzt z\bz}(z\bz)^n}{3}
\left(d\zt\, d\bzt-\frac{5}{9}(z\bz)^{n-1}\frac{z^2 d\bz^2+\bz^2 dz^2}{z\bz}\right)\,.
\eeq

\subsection{$K_{\zt\zt z}K_{\bzt\bzt \bz}$}\label{sec:KztztzKbztbztbz}
This one comes from
\beq
ds^2_{\partial}=d\zt\, d\bzt+2\kappa\left(K_{\zt\zt z}z\, d\zt^2+K_{\bzt\bzt \bz}\bz\, d\bzt^2\right)+dz\, d\bz+O(\kappa^3)\,.
\eeq
Then,
\begin{align}
ds^2=&\left(1-\frac{2}{3}\frac{(n-1)}{1+x}\right)\frac{d\rho^2}{\rho^2}\nonumber\\
&+ \left(1-\frac{2}{3}\frac{(n-1)}{1+x}\right)n^2\frac{\left({1+x}\right)^{\frac{n-1}{n}}}{\rho^{2/n}}dz\, d\bz
\qquad\nonumber\\
&+\left(1+\frac{1}{3}\frac{(n-1)}{1+x}\right)\left[\frac{d\zt\, d\bzt
+2\kappa \left(K_{\zt\zt z}z^n d\zt^2+K_{\bzt\bzt \bz}\bz^n d\bzt^2\right)}{\rho^2}
-4\kappa^2 K_{\zt\zt z}K_{\bzt\bzt \bz}\,d\zt\, d\bzt\right]\nonumber\\
&+ \kappa^2(n-1)ds^2_1+O(\kappa^3)+O\left((n-1)^2\right)\,,
\label{eq:metricKztztzKbztbztbz}
\end{align}
with
\beq
ds_1^2=2K_{\zt\zt z}K_{\bzt\bzt \bz}\left(f^{K0|0}_{\rho\rho}(x)d\rho^2+f^{K0|0}_{z\bz}(x)(z\bz)^{n-1}dz\,d\bz+f^{K0|0}_{\zt\bzt}(x)d\zt\, d\bzt\right)\,.
\eeq
The $\kappa^2 \rho^2$ terms correspond to the Shouten correction to AdS if we were using FG coordinates. This just depends on the boundary geometry and appears at $n=1$.

Discarding the zero mode of \eqref{eq:SSZeroModes}, we find:
\begin{align}
f^{K0|0}_{\rho\rho}(x)=&\frac{2}{3}\left(-5+\frac{1}{1+x}+\log(1+x)\right)\nonumber\\
f^{K0|0}_{\zt\bzt}(x)=&\frac{-1}{3}\left(-16+\frac{1}{1+x}+\log(1+x)\right)-4x\log\left(1+\frac{1}{x}\right)&\nonumber\\
f^{K0|0}_{z\bz}(x)=&\frac{1}{3}\left(-2+\frac{2}{1+x}-\log(1+x)\right)+2x\log\left(1+\frac{1}{x}\right)\,.
\label{eq:solKztztzKbztbztbz}\end{align}
As advertised at the beginning of the section, regularity at the bulk axis now appears to be subtler, as there are no obvious terms with which to absorb these logarithms into $(1+x)^{\frac{n-1}{n}}$ factors.

It helps to notice, however, that the logarithms appear in the same legs as for the two previous cases \eqref{eq:solRzbzzbz}, \eqref{eq:solQztbztzbz}. They can then be cancelled by adding them to the current case, fine-tuned as
\begin{align}
R_{z\bz z\bz}&=-2K_{\zt\zt z}K_{\bzt\bzt \bz}\,,\nonumber\\
Q_{\zt\bzt z\bz}&=2K_{\zt\zt z}K_{\bzt\bzt \bz}\,,
\label{eq:singabs}\end{align}
leading to a regular solution. This solution implies, however, that whenever $K_{\zt\zt z}K_{\bzt\bzt \bz}$ is not zero there will be a part of $R_{z\bz z\bz}$ and $Q_{\zt\bzt z\bz}$ that is \emph{not} accompanied by the $(1+x)^{\frac{n-1}{n}}$ factors, unlike in secs.~\ref{sec:Rzbzzbz} and \ref{sec:Qztbztzbz}. This part, unlike the rest, will have a factor of $(z\bz)^{n-1}$ near the bulk axis.

In practice, this amounts to absorbing the $\log$s in \eqref{eq:solKztztzKbztbztbz} by adding to the ansatz \eqref{eq:metricKztztzKbztbztbz} the following:
\beq
ds^2\rightarrow ds^2+2\kappa^2 K_{\zt\zt z}K_{\bzt\bzt \bz}
\left((z\bz)^n-z\bz(1+x)^{\frac{n-1}{n}}\rho^{2\frac{n-1}{n}}\right)
\left(4\frac{d\zt\,d\bzt}{\rho^2}-2\frac{n^2(1+x)^{\frac{n-1}{n}}\,dz\,d\bz}{\rho^{2/n}}\right)\,.
\eeq
Thus, the $\log$s in \eqref{eq:solKztztzKbztbztbz} are not signalling singular behaviour. We will elaborate on this in sec.~\ref{sec:Splitting}.

The contribution to the stress tensor reads:
\beq
\langle T\rangle=\frac{\ell^3(n-1)}{4\pi G(z\bz)^{2n}}\left(-\kappa^2 K_{\zt\zt z}K_{\bzt\bzt \bz} (z\bz)^n\right)
\left(-2d\zt\, d\bzt+(z\bz)^{n-1}\left(\frac{2}{3}dz\, d\bz+\frac{1}{6}\frac{\bz^2 dz^2 + z^2d\bz^2}{z\bz}\right)\right)\,.
\eeq

\subsection{Summary}
In this section we have written down explicitly smooth gravity duals to all possible squashings of a boundary cone, to the order of Riemann curvature ---with the exception of those involving the trace of the extrinsic curvature, that we set to zero without loss of generality using conformal symmetry.

It is a rather lengthy section because of its exhaustiveness. The cases we have not written down explicitly follow straightforwardly from the ones presented by exchanging $\zt\leftrightarrow\bzt$ and/or $ z\leftrightarrow\bz$.

In the remainder of the paper we will discuss the consequences of these bulk geometries and benchmark them against known features they should reproduce.
\section{Vacuum polarisation}\label{sec:VacPol}
In the preceding section we presented the vacuum polarisation $\langle T \rangle$ induced by a squashed conical singularity in $4D$ holographic conformal field theory. This expectation value of the stress tensor was obtained by conventional holographic techniques \cite{deHaro:2000vlm}, which for us simplify to reverting to Fefferman-Graham-like coordinates and selecting the coefficient of the $\rho^4$ term (see footnote \ref{footnote:FG}).

However, strictly speaking the bulk geometries of this paper do not have well defined Fefferman-Graham expansions, because the boundary metric is singular. It is then necessary to explain in which sense the method used calculate this vacuum polarisation is conventional and valid.

Let us first recall one aspect of the boundary metrics \eqref{eq:AdaptedRepl}. These are obtained by `quotiening' the regular ones in \eqref{eq:AdaptedCoords}, by $z\rightarrow z^n$. The quotient is, away from $z\bz=0$, locally a change of coordinates. As such, it does not alter local properties of the geometry, as the curvature. Hence, the Riemann tensor is regular away from $z\bz=0$. However, the quotient does introduce singular, delta-like, contributions to the curvature at the origin ---the tip of the cone.

Up to these contributions and the multivaluedness discussed at the end of sec.~\ref{sec:FermiCoords}, the boundary geometry behaves regularly, and one can formally develop the Fefferman-Graham expansion and extract a stress tensor. But one needs to bear in mind that the stress tensor calculated in this way ignores contact terms.

As emphasised in the beginning of sec.~\ref{sec:Solutions}, the $\langle T \rangle$ we have extracted should be thought of as belonging to a UV expansion. Indeed, notice that it diverges at the origin as
\beq
\langle T \rangle \sim \frac{n-1}{r^{4n}}\left(\langle T \rangle_0+ \kappa\, r^n\, \langle T \rangle_1+ \kappa^2r^{2n}\langle T \rangle_2+O(\kappa^3)\right)+O\left((n-1)^2\right)\,,
\eeq
where we momentarily reverted to polar coordinates $r^2=z\bz$.

The $\langle T\rangle_k$ contributions we have presented are all local functions of the geometry. Such local probes are characteristic of UV expansions.\footnote{For example, UV divergences of regulated partition functions are local functions of the geometry.} The $r\rightarrow 0$ divergences conform one such expansion. Generically, one also expects there to be finite, non-local dependence on the geometry (as, say, in the ratio of two characteristic lengthscales), but these are missed in the expansion in $\kappa$, because it is around a point. In Fefferman-Graham language, this is concordance with the fact that the range of $\rho$ needs to be small in units of $\kappa$.

For reference, we now collect the full expectation value of the stress tensor:
\begin{align}
\langle T\rangle=\frac{\ell^3(n-1)}{4\pi G(z\bz)^{2n}}&\Bigg\{
\frac{-1}{6}\left(\gamma_{ij}d\sigma^i d\sigma^j+(z\bz)^{n-1}\left(-dz\,d\bz+\frac{\bz^2dz^2+z^2d\bz^2}{z\bz}\right)\right)\nonumber\\
-&\kappa\left[\frac{K_{\zt\zt z}z^n d\zt^2+ K_{\zt\zt \bz}\bz^n d\zt^2}{6}+\textrm{c.c.}\right]\nonumber\\
-&\kappa^2\frac{F_{\zt\bzt z\bz}}{2}
(z\bz)^{n-1}(z\,d\bz-\bz\, dz)(\zt\,d\bzt-\bzt\, d\zt)
\nonumber\\
+&\kappa^2\left[\frac{Q_{\zt\bzt zz}\,z^{2n}}{9}
\left(d\zt\, d\bzt+(z\bz)^{n-1}\left(-4dz\,d\bz+2\frac{z^2 d\bz^2+3\bz^2 dz^2}{z\bz}\right)\right)
+\textrm{c.c.}\right]
\nonumber\\
+&\kappa^2\left[\frac{K_{\zt\zt z}K_{\bzt\bzt z} z^{2n}}{9}
\left(-5d\zt\, d\bzt+(z\bz)^{n-1}\left(17dz\,d\bz-\frac{7z^2 d\bz^2+30\bz^2 dz^2}{z\bz}\right)\right)
+\textrm{c.c.}\right]
\nonumber\\
-&\kappa^2\left[\frac{Q_{\zt\zt z\bz} (z \bz)^n d\zt^2}{3}
+\textrm{c.c.}\right]
\nonumber\\
-&\kappa^2\left[ \frac{K_{\zt\zt z,\bzt} z^n d\zt}{6}
(z\bz)^{n-1}\left(\frac{3}{2}z\, d\bz-2\bz\, dz\right)
+\zt\leftrightarrow\bzt+\textrm{c.c.}\right]
\nonumber\\
+&\kappa^2\left[\frac{ R_{\zt z z\bz} z^n d\zt}{36}
(z\bz)^{n-1}\left(8\bz\, dz+13 z\, d\bz\right)
+\zt\leftrightarrow\bzt+\textrm{c.c.}\right]
\nonumber\\
-&\kappa^2
\frac{5{\cal R}_{\zt\bzt \zt\bzt}}{27}(z\bz)^{2(n-1)}(z^2 d\bz^2+\bz^2 dz^2)
\nonumber\\
+&\kappa^2\frac{R_{z \bz z \bz}(z\bz)^n}{3}
\left(2d\zt\, d\bzt+(z\bz)^{n-1}\left(-dz\, d\bz+\frac{4}{9}\frac{\bz^2 dz^2+z^2d\bz^2}{z\bz}\right)
\right)
\nonumber\\
-&\kappa^2\frac{2 Q_{\zt\bzt z\bz}(z\bz)^n}{3}
\left(d\zt\, d\bzt-\frac{5}{9}(z\bz)^{n-1}\frac{z^2 d\bz^2+\bz^2 dz^2}{z\bz}\right)
\nonumber\\
-&\kappa^2 \left(K_{\zt\zt z}K_{\bzt\bzt \bz}+\zt\leftrightarrow\bzt\right) (z\bz)^n
\left(-2d\zt\, d\bzt+(z\bz)^{n-1}\left(\frac{2}{3}dz\, d\bz+\frac{1}{6}\frac{\bz^2 dz^2 + z^2d\bz^2}{z\bz}\right)\right)
\nonumber\\
+&O(\kappa^3)\Bigg\}+O\left((n-1)^2\right)\,,
\label{eq:FullStressTensor}\end{align}
which is traceless. We have left implicit that the metric on the surface $\gamma_{ij}$ may have intrinsic curvature, and that the extrinsic curvature may have dependence on the position on the surface. This expression may be seen as encoding a number of response coefficients of the field theory to the presence of the squashed cone.

Not having fixed the conformal frame completely, we can perform the sanity check that this stress tensor transforms covariantly under Weyl rescalings.\footnote{The conformal anomaly gets activated at $O(\kappa^4)$ ($O(\kappa^2)$ in the contact terms), so it plays no role here.} Consider the case in which only the following squashing is turned on: $R_{z\bz z\bz}=2Q_{\zt\bzt z\bz}=\frac{1}{2}$. This is conformally flat:
\beq
ds^2_\partial=\left[1+\kappa^2 z\bz\right]\left(d\zt\, d\bzt +dz\,d\bz\right)+O(\kappa^3)\,.
\label{eq:ConfFlat}\eeq
The stress tensor induced by the introduction of the cone on \eqref{eq:ConfFlat} (by $z\rightarrow z^n$) can be read from \eqref{eq:FullStressTensor}. Its dependence on $\kappa$ also displays conformal flatness:
\begin{align}
\langle T\rangle=\left[\frac{1}{1+\kappa^2 (z\bz)^n}\right]\frac{\ell^3(n-1)}{4\pi G(z\bz)^{2n}}
\frac{-1}{6}\left(\gamma_{ij}d\sigma^i d\sigma^j+(z\bz)^{n-1}\left(-dz\,d\bz+\frac{\bz^2dz^2+z^2d\bz^2}{z\bz}\right)\right)\quad&\nonumber\\
+O(\kappa^3)
+O\left((n-1)^2\right)\,,&
\end{align}
providing a check of the good conformal covariance properties of \eqref{eq:FullStressTensor}. Similarly, the other two locally conformally flat cases, ${\cal R}_{\zt\bzt\zt\bzt}=-R_{z\bz z\bz}$ and $Q_{\zt\bzt z z}\neq 0$, can also be seen to follow from the one without bending, $\kappa=0$.

\section{Logarithmic divergences of holographic R\'enyi entropy}\label{sec:Renyi}
The metrics dual to squashed cones reproduce the results of \cite{Dong:2016wcf} regarding logarithmic divergences of R\'enyi entropy for holographic CFTs.

Entanglement entropy is known to be UV divergent in field theory. This divergence is due to correlations across the entangling surface between infinitely many short distance degrees of freedom. It is therefore localised around the entangling surface. Taming it with a short distance cutoff, it reads, for the vacuum of a 4D CFT \cite{Solodukhin:2008dh}:
\beq
S=\frac{\textrm{Area}}{\epsilon^2}+\left(\frac{a}{2\pi}\int {\cal R}\,\sqrt{\gamma}\, d^2\sigma+\frac{c}{2\pi}\int\left(K_{\{ij\}a}K^{\{ij\}a}
-W_{ij}{}^{ij}\right)\sqrt{\gamma}\,d^2\sigma \right)\log\epsilon+\dots\,.
\eeq
where the area and the integrals are on the entangling surface. ${\cal R}$ is the Ricci scalar of the induced metric $\gamma_{ij}$; $W_{ij}{}^{ij}$ is the contraction of the projection of the Weyl tensor on the surface; and $K_{\{ij\}a}K^{\{ij\}a}$ is the contraction of the square of the traceless part of the extrinsic curvature. The last two are conformal invariant.

While the coefficient of the area term is sensitive to the choice of cutoff, the logarithmic divergence is not. It therefore has a physical character. $a$ and $c$ are the central charges ---the logarithmic divergence can be derived from the conformal anomaly, when the latter is written as the logarithmic divergence of the regulated effective action \cite{Solodukhin:2008dh}.

R\'enyi entropies are conjectured to have a similar UV behavior \cite{Fursaev:2012mp}:
\beq
S_n=\dots+\left(\frac{f_a(n)}{2\pi}\int {\cal R}\,\sqrt{\gamma}\, d^2\sigma+\frac{f_b(n)}{2\pi}\int K_{\{ij\}a}K^{\{ij\}a}\sqrt{\gamma}\,d^2\sigma-\frac{f_c(n)}{2\pi}\int W_{ij}{}^{ij}\sqrt{\gamma}\,d^2\sigma \right)\log\epsilon+\dots\,.
\eeq
There are known relations between $f_a(n)$ and $f_c(n)$ \cite{Perlmutter:2013gua, Lewkowycz:2014jia}, but less is known about $f_b(n)$, apart from $f_b(1)=c$. Free field theory results prompted the conjecture that $f_b(n)=f_c(n)$ \cite{Lee:2014xwa}. However, this relation fails for holographic theories \cite{Dong:2016wcf}. The method of \cite{Dong:2016fnf} applied to our metrics for duals to squashed cones reproduces this failing. This method builds on \cite{Hung:2011nu} and \cite{Lewkowycz:2013nqa} to argue that a certain derivative of R\'enyi entropy with respect to the index $n$ is given, for theories holographically dual to General Relativity, by the area of the bulk axis:
\beq
n^2\partial_{n}\left(\frac{n-1}{n}S_n\right)=\frac{\textrm{Area}(\textrm{axis})}{4G}\,.
\label{eq:RenyiFromArea}\eeq
This formula is analogous to the fact that, in thermodynamics, the thermal derivative of the free energy is the entropy.\footnote{Here, R\'enyi entropy is analogous to thermodynamic free energy, not thermodynamic entropy.} In General Relativity, this entropy is an area.

We thus need to calculate the area of the bulk axis (at $z\bz\rightarrow 0$) of the metrics of sec.~\ref{sec:Solutions}. Given that these metrics are precise to $O(\kappa^2)$ and $O(n-1)$, we can extract the curvature contributions to the R\'enyi entropy to first order in $(n-1)$.

Given eq.~\eqref{eq:DualCone}, $O(\kappa^2)$ contributions to the area of the axis can come only from $g_{\zt\bzt}$ and $g_{\rho\rho}$ at $z\bz=0$. These are non-zero only for the $0|0$ cases studied in secs.~\ref{sec:Rztbztztbzt}--\ref{sec:KztztzKbztbztbz}. There is a class of such contributions that remains finite in the $n\rightarrow 1$ limit ---the Shouten terms. In this limit, $S_n$ in \eqref{eq:RenyiFromArea} becomes just entanglement entropy. Therefore, the Shouten terms determine the shape dependence of entanglement entropy \cite{Graham:1999pm, Schwimmer:2008yh}.

At $O(n-1)$, there is an interplay between the Shouten terms multiplied by the $\frac{(n-1)}{1+x}$ factors of \eqref{eq:DualCone}, that do not vanish at the axis; and from the fact that the functions in \eqref{eq:solRztbztztbzt}, \eqref{eq:solRzbzzbz}, \eqref{eq:solQztbztzbz} and \eqref{eq:solKztztzKbztbztbz} do not vanish at $x\rightarrow 0$. Collecting all terms, one gets the following area density at the axis:
\begin{align}
a(\textrm{axis})
=\Bigg(&
\frac{1}{\rho^3}
+\frac{{\cal R}_{\zt\bzt\zt\bzt}}{\rho}\left(\frac{4}{3}-\frac{20}{27}(n-1)\right)
+\frac{Q_{\zt\bzt z\bz}}{\rho}\left(\frac{4}{3}-\frac{68}{27}(n-1)\right)
\nonumber\\
&
+\frac{R_{z\bz z\bz}}{\rho}\left(-\frac{2}{3}+\frac{34}{27}(n-1)\right)
+\frac{K_{\zt\zt z} K_{\bzt\bzt \bz}+K_{\zt\zt \bz} K_{\bzt\bzt z}}{\rho}\left(-4+\frac{22}{3}(n-1)\right)\Bigg)\sqrt{\gamma}\nonumber\\
&
+O(\kappa^3)+O\left((n-1)^2\right)\,,
\end{align}
where we include the $K_{\zt\zt \bz}K_{\bzt\bzt z}$ contribution that was left implicit in sec.~\ref{sec:Solutions}. Its presence follows from covariance.
From this, a `R\'enyi entropy density' follows by integrating in \eqref{eq:RenyiFromArea}. This density is to be integrated in the three directions that span the axis: $\rho$, $\zt$ and $\bzt$. Performing the integral in $\rho$ from a cutoff $\rho=\epsilon$ inwards, and using \eqref{eq:PWeyl2}, we get
\begin{align}
S_n=\frac{1}{16G}\frac{1}{\epsilon^2}\int\sqrt{\gamma}\,d^2\sigma
+\frac{\kappa^2}{16G}\Bigg[&\left(1-\frac{1}{2}(n-1)\right)\int{\cal R}\,\sqrt{\gamma}\,d^2\sigma
\nonumber\\
&+\left(1-\frac{11}{12}(n-1)\right)\int  K_{\{ij\}a}K^{\{ij\}a}\,\sqrt{\gamma}\,d^2\sigma
\nonumber\\
&-\left(1-\frac{17}{18}(n-1)\right)\int W_{ij}{}^{ij}\,\sqrt{\gamma}\,d^2\sigma\Bigg]\log\epsilon
\nonumber\\
&+O(\kappa^3)+O\left((n-1)^2\right)\,.
\label{eq:RenyiLog}\end{align}
Notice that the area divergence does not have a correction in $(n-1)$. In fact, it is easy to argue that in holography such area divergence does not have dependence in $n$, as this divergence follows from the leading term at small $\rho$ of $g_{\rho\rho}$ and $g_{\zt\bzt}$, and these do not change under $z\rightarrow z^n$.

Upon restoring units of $\ell$ and using that $a=c=\pi\ell^3/8G$, \eqref{eq:RenyiLog} reproduces \cite{Dong:2016wcf}. For this theory, indeed, $f_b\neq f_c$.

\section{Splitting problem and singularity resolution in the bulk}\label{sec:Splitting}
This section discusses the impact of the analysis of sec.~\ref{sec:KztztzKbztbztbz} on the holographic entanglement entropy formula of higher-derivative theories of gravity. We will conclude, in precise agreement with \cite{Miao:2015iba}, that the `splitting problem' has a non-minimal solution in a class of theories, resulting in a slightly different entropy formula from previously anticipated in \cite{Dong:2013qoa, Camps:2013zua}. This difference is visible only beyond curvature squared interactions, and does not impact the entropy of Lovelock nor $f(R)$ interactions. We include a brief but self-contained description of the entropy formula for higher-derivative gravity and its splitting problem.

The application of generalized entropy to higher-derivative theories of gravity results in a new holographic entanglement entropy formula. For the class of theories with a lagrangian depending on the Riemann tensor but not on its derivatives,\footnote{For continuity with the rest of this paper we consider five bulk dimensions, although the applicability is more general.}
\beq
I=\int  L(\textrm{Riem})\,\sqrt{g}\,d^5 x+\textrm{Boundary terms},
\label{eq:HigherDers}\eeq
this formula is
\beq
S=2\pi\int\left\{\frac{\partial L}{\partial R_{z\bz z\bz}}+8\sum_\alpha
\left(\frac{\partial^2 L}{\partial R_{z i z j} \partial R_{\bz k \bz l}}\right)_{\alpha}\frac{K_{ijz}K_{kl\bz}}{q_\alpha+1}\right\}\sqrt{\gamma}\,d^3x\,.
\label{eq:WaldP}\eeq
Here $R_{z\bz z\bz}$, $R_{zizj}$, $K_{ijz}$ and their complex conjugates refer to an expansion of the type \eqref{eq:AdaptedCoords} around the bulk entangling surface, at $z\bz\rightarrow 0$. This surface is what we called the `bulk axis' in previous sections. In contrast to the use of the expansion \eqref{eq:AdaptedCoords} in sec.~\ref{sec:FermiCoords}, $i$ and $j$ now run over three values, that in the coordinates of that section would be $\zt$, $\bzt$ and $\rho$.

The first term in this holographic entropy formula is Wald entropy \cite{Wald:1993nt}, and the second one can be thought of as a correction to it. Wald entropy was constructed on bifurcation surfaces of event horizons, which necessarily have vanishing extrinsic curvature. The entropy of \eqref{eq:WaldP} applies also to situations in which the extrinsic curvature may be non-zero.

To explain the meaning of the sum in $\alpha$ in \eqref{eq:WaldP} we need to discuss some details of the application of generalized entropy to the class of theories \eqref{eq:HigherDers}. This application involves evaluating actions of bulk geometries\footnote{rather, actions of `analytic continuations of geometries' (see comments after eq.~\ref{eq:SurfaceGauge}).} that regulate conical singularities. The prescription of \cite{Dong:2013qoa, Camps:2013zua} for the sum in $\alpha$ assumes a `minimal' regulation, of the type discussed below eq.~\eqref{eq:ConeCorr}. Here minimal means that, in an expansion around the axis (at $z\bz=0$), the metric of the regulated cone takes the form:
\begin{align}
ds^2=&\left(\gamma_{ij}+\left[2\kappa\, K_{ijz}{} z^n+\kappa^2Q_{ijzz}z^{2n}+\kappa^2Q_{ijz\bz}z\bz+\textrm{c.c.}\right]\right)d\sigma^i d\sigma^j+2\kappa\, A_{iz\bz} (\bz\, dz-z\,d\bz)\,d\sigma^i \nonumber\\
&-\frac{4}{3}\kappa^2\left[R_{izz\bz}z^n-\textrm{c.c.}\right] (\bz\, dz-z\, d\bz) d\sigma^i+\left(1+2\kappa^2R_{z\bz z\bz}z\bz\right)dz\,d\bz+O(\kappa^3)\,.
\label{eq:MinimalRegulation}
\end{align}

This follows from taking \eqref{eq:AdaptedCoords} and promoting any holomorphic factors of $z$ and $dz$ that are not paired with antiholomorphic ones to $z^n$ and $d(z^n)=n z^{n-1} dz$, respectively. This achieves a replica symmetric metric \eqref{eq:MinimalRegulation} that is regular at the axis for integer $n$, and for which the exponents in $z$ differ minimally from the ones before replicating \eqref{eq:AdaptedCoords}.

The entropy following from the evaluation of the action of such cone-regulating geometries involves integrals of the type
\beq
\lim_{n\rightarrow 1}\partial_n\int_0^{\infty}(n-1)^2\left(\frac{r^{n}}{r^2}\right)^2r^{2q_\alpha(n-1)}e^{-r^2}r\,dr
=\frac{1}{2}\frac{1}{q_{\alpha}+1}\,,
\label{eq:Integral}\eeq
where we used polar coordinates $r=\sqrt{z\bz}$. The role of the exponential function is to localise around the axis, and this function could be replaced without change in the rhs by any other regulating function, interpolating smoothly between $1$ at the origin and $0$ at infinity ---\emph{e.g.}, $\frac{1}{1+r^2}$.

The integral in eq.~\eqref{eq:Integral} is dominated by a logarithmic divergence at the lower end as $n\rightarrow 1$. That explains the independence from the regulating function. The outcome is sensitive to the details of the expansion of the geometry around the axis, that are encoded in the $q_{\alpha}$ in \eqref{eq:Integral}.  $q_\alpha$ parametrises $n-$dependence in the power of $r$ in the integrand, reflecting $n-$dependent exponents of $z$ and $\bz$ in the geometry \eqref{eq:MinimalRegulation}.

Terms in the expansion \eqref{eq:MinimalRegulation} that are accompanied by different powers of $z$ and $\bz$ contribute differently to $q_\alpha$. That is what the sum in $\alpha$ in \eqref{eq:WaldP} captures. In this formula we need to decompose the second derivative of the lagrangian in monomials of the curvature, that  $\alpha$ labels. These monomials are, however, not in the background Riemann tensor, as may appear natural for $L(\textrm{Riem})$. Rather, its constituents are the quantities appearing \eqref{eq:MinimalRegulation}, in terms of which one can write the Riemann tensor \eqref{eq:GC}: ${\cal R}_{ijkl}$, $K_{ijz}$, $Q_{ijzz}$, $Q_{ijz\bz}$, $R_{izz\bz}$, $F_{ijz\bz}$ and $R_{z\bz z\bz}$ (and complex conjugates).  Each monomial $\alpha$ is then assigned a value of $q_{\alpha}$, and the sum is performed with the $\frac{1}{q_{\alpha}+1}$ weight. Constituents of $\alpha$ contribute additively to $q_{\alpha}$ with a weight that depends on the exponent of $z$ and $\bz$ that they are accompanied by around the axis of the regulated cone. For the regulation of \eqref{eq:MinimalRegulation}, $q_{\alpha}$ is contributed $1/2$ for each $K_{ijz}$ and $R_{izz\bz}$, $1$ for $Q_{ijzz}$, and $0$ otherwise.

Note, however, that exchanging, \emph{e.g.}, the $z\bz$ factor accompanying $R_{z\bz z\bz}$ in \eqref{eq:MinimalRegulation} for $(z\bz)^n$  would also achieve a regular replica symmetric metric, although with a different weight of this term in \eqref{eq:WaldP}. Now, $R_{z\bz z\bz}$  would contribute $1$ to $q_{\alpha}$, instead of $0$. The obvious such ambiguities are in terms with a $z\bz$ pair in their indices: $Q_{ijz\bz}$, $F_{ijz\bz}$, $R_{izz\bz}$ and $R_{z\bz z\bz}$; although there may be more \cite{Camps:2014voa}. These ambiguities have been called `the splitting problem' \cite{Miao:2015iba}.

A lesson that follows from the analysis in sec.~\ref{sec:KztztzKbztbztbz} is that, in General Relativity, the expansion around the bulk axis does \emph{not} take the form of eq.~\eqref{eq:MinimalRegulation}. Rather, the $Q_{\zt\bzt z\bz}$ and $R_{z\bz z\bz}$ terms look:\footnote{The new terms compared to eq.~\eqref{eq:singabs} follow from covariance. Strictly speaking, eq.~\eqref{eq:GRsplitting} is not dimensionally correct. To avoid clutter, we omit the factors of $\rho^{2\frac{n-1}{n}}$ that would render it so. These are finite at the bulk axis.}
\begin{align}
ds^2=\dots
&+4\kappa^2\left[Q^{\prime}_{\zt\bzt z\bz}\,z\bz+2\left(K_{\zt\zt z}K_{\bzt\bzt \bz}+K_{\zt\zt \bz}K_{\bzt\bzt z}\right)(z\bz)^n\right]d\zt\, d\bzt
\nonumber\\
&
+2\kappa^2\left[R^{\prime}_{z\bz z\bz}\,z\bz-2\left(K_{\zt\zt z}K_{\bzt\bzt \bz}+K_{\zt\zt \bz}K_{\bzt\bzt z}\right)(z\bz)^n\right]dz\, d\bz
+\dots+O(\kappa^3)\,.
\label{eq:GRsplitting}
\end{align}
with
\begin{align}
Q^{\prime}_{\zt\bzt z\bz}&=Q_{\zt\bzt z\bz}-2\left(K_{\zt\zt z}K_{\bzt\bzt \bz}+K_{\zt\zt \bz}K_{\bzt\bzt z}\right)\label{eq:Qprime}\\
R^{\prime}_{z\bz z\bz}&=R_{z\bz z\bz}+2\left(K_{\zt\zt z}K_{\bzt\bzt \bz}+K_{\zt\zt \bz}K_{\bzt\bzt z}\right)\,.\label{eq:Rprime}
\end{align}
Notice that upon taking $n\rightarrow 1$ the extrinsic curvature contributions cancel and we recover \eqref{eq:AdaptedCoords}.

Let us for a moment discuss what is the $\sigma^i-$covariant version of eqs.~\eqref{eq:Qprime} and \eqref{eq:Rprime}. Recall that, in this section, $\sigma^i$ encompasses $\zt$, $\bzt$ and $\rho$. This is irrelevant for the second term, that can be written
\beq
R^{\prime}_{z\bz z\bz}=R_{z\bz z\bz}+\frac{1}{2}K_{ijz}K^{ij}{}_{\bz}\,,
\eeq
because for the configurations we studied in sec.~\ref{sec:Solutions}, $K_{\rho i z}=0$. Notice also that in this section $K_{ijz}$ can not have a trace ---as the Ryu-Takayanagi surface is a minimal surface.

For  $Q^{\prime}$ a similar argument implies that the covariantisation should read:
\beq
Q^{\prime}_{ijz\bz}=Q_{ij z\bz}-K_{ikz}\gamma^{kl}K_{jl\bz}
-K_{ik\bz}\gamma^{kl}K_{jlz}\,.
\eeq
Equivalent expressions for the analogs of $Q^{\prime}$ and $R^{\prime}$ were found in \cite{Miao:2015iba} by solving the Einstein equations around the bulk axis.\footnote{In comparison to that reference, we have used conformal symmetry to eliminate the trace of the extrinsic curvature.}

Since the factors of $(z\bz)^n$ in \eqref{eq:GRsplitting} are different from those in \eqref{eq:MinimalRegulation}, we conclude that the splitting problem has a non-minimal solution in General Relativity. This translates into the $\alpha$ sum of \eqref{eq:WaldP} meaning something different than it would in the minimal case of \eqref{eq:MinimalRegulation}. Now, $\alpha$ labels monomials in terms of  ${\cal R}_{ijkl}$, $K_{ijz}$, $Q_{ijzz}$, $Q^{\prime}_{ijz\bz}$, $R_{izz\bz}$, $F_{ijz\bz}$ and $R^{\prime}_{z\bz z\bz}$; instead of $Q_{ij z\bz}$ and $R_{z\bz z\bz}$. Explicit factors of $K_{ijz}$ and $R_{izz\bz}$ still contribute $1/2$ to $q_{\alpha}$; $Q_{ijzz}$ contributes $1$; and the rest, including $Q^{\prime}_{\zt\bzt z\bz}$  and $R^{\prime}_{z\bz z\bz}$, contribute $0$.

This may appear irrelevant, since the lagrangian of GR has a vanishing second derivative in the Riemann and therefore no splitting problem; its entropy is just the area. However, this splitting does have consequences for the entropy formula of theories that contain perturbative higher-derivative corrections to General Relativity. For these corrections there is a splitting problem, and the splitting is fixed by the leading result ---the GR one we just discussed.

This affects $\textrm{Riem}^k$ interactions for $k\geq3$, but does not have consequences for Lovelock interactions, because in those $\frac{\partial^2 L}{\partial R_{z i z j} \partial R_{\bz k \bz l}}$ does not depend on $Q_{\zt\bzt z\bz}$  nor $R_{z\bz z\bz}$;\footnote{This comment is non-trivial only when the bulk dimension is $D\geq6$, when the Lovelock term of order $>2$, and thus with a splitting ambiguity, becomes non-trivial.} or for $f(R)$ interactions, for which the second derivative vanishes identically.
%

As an illustration of the consequences of this resolution of the splitting problem, consider
\beq
L=-\frac{1}{16\pi G}R-\frac{\lambda}{2^6}\, R_{\mu\nu}{}^{\rho\sigma}R_{\rho\sigma}{}^{\tau\omega}R_{\tau\omega}{}^{\mu\nu}+O(\lambda^2)\,.
\eeq
For this theory (leaving implicit the $O(\lambda^2)$):
\beq
\frac{\partial^2 L}{\partial R_{z i z j} \partial R_{\bz k \bz l}}=-\frac{3}{8}\lambda\left(\delta_{il}R_{zj\bz k}+\delta_{jk}R_{zi\bz l}\right)\,,
\eeq
where the $3$ is a symmetry factor and there is a factor of $2^3$ from three $g^{z\bz}$. From here we get
\beq
\frac{\partial^2 L}{\partial R_{z i z j} \partial R_{\bz k \bz l}}K^{ij}{}_{z} K^{kl}{}_{\bz}=-3\lambda\, R_{z\zt\bz \bzt}K_{\zt\zt z}K_{\bzt\bzt \bz}+{\zt\leftrightarrow \bzt}\,,
\eeq
and with the splitting of \eqref{eq:GRsplitting}:
\begin{align}
\sum_{\alpha}\left(\frac{\partial^2 L}{\partial R_{z i z j} \partial R_{\bz k \bz l}}\right)_{\alpha}\frac{K_{ijz} K_{ij\bz}}{q_{\alpha}+1}
&=-3\lambda\sum_{\alpha} \left(R_{z\zt\bz \bzt}\right)_{\alpha}\frac{K_{\zt\zt z}K_{\bzt\bzt \bz}}{q_{\alpha}+1}+{\zt\leftrightarrow \bzt}\nonumber\\
&=-3\lambda\sum_{\alpha} \left(\frac{1}{2}F_{\zt\bzt z \bz}-Q_{\zt \bzt z\bz}+2K_{\zt\zt \bz}K_{\bzt\bzt z}\right)_{\alpha}\frac{K_{\zt\zt z}K_{\bzt\bzt \bz}}{q_{\alpha}+1}+{\zt\leftrightarrow \bzt}\nonumber\\
&=-3\lambda\sum_{\alpha} \left(\frac{1}{2}F_{\zt\bzt z \bz}-Q^{\prime}_{\zt \bzt z\bz}-2K_{\zt\zt z}K_{\bzt\bzt\bz}\right)_{\alpha}\frac{K_{\zt\zt z}K_{\bzt\bzt \bz}}{q_{\alpha}+1}+{\zt\leftrightarrow \bzt}\nonumber\\
&=-3\lambda \left(\frac{1}{2}F_{\zt\bzt z \bz}-Q^{\prime}_{\zt \bzt z\bz}-\frac{2}{2}K_{\zt\zt z}K_{\bzt\bzt\bz}\right) K_{\zt\zt z}K_{\bzt\bzt \bz}+{\zt\leftrightarrow \bzt}\nonumber\\
&=-3\lambda \left(\frac{1}{2}F_{\zt\bzt z \bz}-Q_{\zt \bzt z\bz}+K_{\zt\zt z}K_{\bzt\bzt\bz}+2K_{\zt\zt \bz}K_{\bzt\bzt z}\right) K_{\zt\zt z}K_{\bzt\bzt \bz}+{\zt\leftrightarrow \bzt}\nonumber\\
&=-3\lambda \left(R_{z\zt\bz \bzt}+K_{\zt\zt z}K_{\bzt\bzt\bz}\right) K_{\zt\zt z}K_{\bzt\bzt \bz}+{\zt\leftrightarrow \bzt}+O(\lambda^2)\,,
\end{align}
in agreement with \cite{Miao:2015iba}. This could be simplified further using the $0$th order background eoms (Einstein's).\footnote{We have used the simplification $K_{\zt\bzt z}=0$; the Ryu-Takayangi surface is extremal in General Relativity ($\lambda=0$).} For the second line we have used eq.~\eqref{eq:GC}, and for the third, \eqref{eq:Qprime}. In the fourth line we summed over $\alpha$ with the splitting we have discussed, and in the fifth and sixth we have used eqs.~\eqref{eq:Qprime} and \eqref{eq:GC} again. For the `minimal splitting', we would perform the sum in $\alpha$ directly from the second line, getting in the end $-3\lambda \left(R_{z\zt\bz \bzt}-K_{\zt\zt\bz}K_{\bzt\bzt z}\right) K_{\zt\zt z}K_{\bzt\bzt \bz}+{\zt\leftrightarrow \bzt}$, for which the $K^4$ term is a different tensor structure altogether.

\section{Outlook}\label{sec:Conclusions}
This paper has described the regular GR duals to CFTs on squashed cones. These metrics show how bulk gravity regulates a conical singularity in the boundary. We have worked to first order in the strength of the cone $(n-1)$, and to second order in $\kappa$, parametrising an ultralocal expansion around the cone. Going to second order allows sensitivity to Riemann curvature ---although only within a UV expansion.

A quantity that follows from these geometries is (the UV expansion of) the vacuum polarisation in the presence of these cones \eqref{eq:FullStressTensor}, up to contact terms. This stress tensor bears some resemblance to those of fluids/gravity, and one can interpret its many coefficients as response coefficients to the squashing of the cone. The number of such second-order coefficients is large compared to \cite{Bhattacharyya:2008mz} because the entangling surface breaks $O(4)$ symmetry to two planes (parallel and transverse). Since we did not take full advantage of conformal symmetry, three of the contributions to \eqref{eq:FullStressTensor} can be generated via covariance under conformal transformations, as we saw in sec.~\ref{sec:VacPol}. For closure, it would be interesting to write down the missing contact terms, that may be interpreted as defect degrees of freedom.

Our setup should not be confused with that of holographic entanglement entropy across surfaces with singular shapes \cite{Bueno:2015xda}. In that setup, the background boundary metric is regular.

The metric of a conical singularity at the boundary is simple both in complex coordinates \eqref{eq:ConeComplex} and in hyperbolic ones \eqref{eq:ThH}. \eqref{eq:ThH} has a simple gravity dual for all values of $n$, eq.~\eqref{eq:HypBH}, while for \eqref{eq:ConeComplex} we have worked only to leading order in $(n-1)$, \eqref{eq:DualCone}. It is natural to suspect that there should be a simple gravity dual to \eqref{eq:ConeComplex} for all values of $n$, and a correspondingly simple generalisation of our results non-linearly in $n$. Such generalisation would be applicable, \emph{e.g.}, to negativity as the $n\rightarrow 1/2$ limit \cite{Perlmutter:2015vma}. We plan to investigate this elsewhere. It should also be possible to generalise the results of this paper to other bulk dimensions, and to other theories of gravity.

The detailed mechanism by which the bulk regulates the boundary cone is in agreement with \cite{Miao:2015iba}. There, this structure was derived by solving the finite part of Einstein's equations around the Ryu-Takayanagi surface (the infinite part gives that the surface is minimal \cite{Lewkowycz:2013nqa}). That suggests that the addition of matter may change the detailed regulation, and therefore the solution of the splitting problem we presented in sec.~\ref{sec:Splitting}. It may be interesting to explore this possibility, and whether it impacts the $\log$ divergence of R\'enyi entropy we discussed in sec.~\ref{sec:Renyi} ---perhaps there is after all a gravity dual for which $f_b(n)=f_c(n)$ is realised.

\section*{Acknowledgements}
It is a pleasure to acknowledge conversations with Garrett Goon, Shahar Hadar, and David Tong; and correspondence with Xi Dong. Work supported by the ERC grant agreement STG
279943, `Strongly Coupled Systems'.

\end{document}